\documentclass[conference]{IEEEtran}

\usepackage{cite}
\usepackage{amsmath,amssymb,amsfonts}
\usepackage{algorithmic}
\usepackage{graphicx}
\usepackage{textcomp}
\usepackage{xcolor}
\def\BibTeX{{\rm B\kern-.05em{\sc i\kern-.025em b}\kern-.08em
    T\kern-.1667em\lower.7ex\hbox{E}\kern-.125emX}}

\usepackage[linesnumbered,ruled,vlined,commentsnumbered]{algorithm2e}
\usepackage{epsfig,endnotes,array}
\usepackage{shortcuts,pifont}
\usepackage{booktabs}
\usepackage{graphicx}
\usepackage{subcaption}
\usepackage{color,caption,ulem}
\definecolor{lightgray}{rgb}{0.95, 0.95, 0.95}
\definecolor{darkgray}{rgb}{0.4, 0.4, 0.4}
\definecolor{editorGray}{rgb}{0.95, 0.95, 0.95}
\definecolor{editorOcher}{rgb}{1, 0.5, 0} 
\definecolor{editorGreen}{rgb}{0, 0.5, 0} 
\definecolor{orange}{rgb}{1,0.45,0.13}		
\definecolor{olive}{rgb}{0.17,0.59,0.20}
\definecolor{brown}{rgb}{0.69,0.31,0.31}
\definecolor{purple}{rgb}{0.38,0.18,0.81}
\definecolor{lightblue}{rgb}{0.1,0.57,0.7}
\definecolor{lightred}{rgb}{1,0.4,0.5}
\usepackage{xspace}
\usepackage{upquote}
\usepackage{listings}
\usepackage{multirow}
\usepackage{float}
\restylefloat{table}
\PassOptionsToPackage{hyphens}{url}\usepackage{hyperref}
\hypersetup{
	colorlinks = true,
	linkcolor = blue,
	anchorcolor = blue,
	citecolor = blue,
	filecolor = blue,
	urlcolor = blue
}
\usepackage{xurl}
\usepackage[htt]{hyphenat}

\lstdefinelanguage{CSS}{
	keywords={color,background-image:,margin,padding,font,weight,display,position,top,left,right,bottom,list,style,border,size,white,space,min,width, transition:, transform:, transition-property, transition-duration, transition-timing-function},	
	sensitive=true,
	morecomment=[l]{//},
	morecomment=[s]{/*}{*/},
	morestring=[b]',
	morestring=[b]",
	alsoletter={:},
	alsodigit={-}
}

\lstdefinelanguage{JavaScript}{
	morekeywords={typeof, new, true, false, catch, function, return, null, catch, switch, var, if, in, while, do, else, case, break},
	morecomment=[s]{/*}{*/},
	morecomment=[l]//,
	morestring=[b]",
	morestring=[b]'
}

\lstdefinelanguage{HTML5}{
	language=html,
	sensitive=true,	
	alsoletter={<>=-},	
	morecomment=[s]{<!-}{-->},
	tag=[s],
	otherkeywords={
		>,
		<!DOCTYPE,
		</html, <html, <head, <title, </title, <style, </style, <link, </head, <meta, />,
		</body, <body,
		</div, <div, </div>, 
		</p, <p, </p>,
		</script, <script,
		<canvas, /canvas>, <svg, <rect, <animateTransform, </rect>, </svg>, <video, <source, <button, </button, <iframe, </iframe, </video>, <img, </img, <header, </header, <article, </article
	},
	ndkeywords={
		=,
		charset=, src=, id=, width=, height=, style=, type=, rel=, href=,
		fill=, attributeName=, begin=, dur=, from=, to=, poster=, controls=, x=, y=, repeatCount=, xlink:href=,
		margin:, padding:, background-image:, border:, top:, left:, position:, width:, height:, margin-top:, margin-bottom:, font-size:, line-height:,
		transform:, -moz-transform:, -webkit-transform:,
		animation:, -webkit-animation:,
		transition:,  transition-duration:, transition-property:, transition-timing-function:,
	}
}

\lstdefinestyle{http} {%
	basicstyle={\footnotesize\ttfamily},   
	frame=b,
	identifierstyle=\color{black},
	keywordstyle=\color{blue}\bfseries,
	ndkeywordstyle=\color{editorGreen}\bfseries,
	stringstyle=\color{editorOcher}\ttfamily,
	commentstyle=\color{brown}\ttfamily,
	language=HTML5,
	alsodigit={.:;},	
	tabsize=1,
	showtabs=false,
	showspaces=false,
	showstringspaces=false,
	extendedchars=true,
	breaklines=true,
	literate=%
	{Ö}{{\"O}}1
	{Ä}{{\"A}}1
	{Ü}{{\"U}}1
	{ß}{{\ss}}1
	{ü}{{\"u}}1
	{ä}{{\"a}}1
	{ö}{{\"o}}1
}

\lstdefinestyle{htmlcssjs} {
	basicstyle={\footnotesize\ttfamily},   
	frame=b,
	xleftmargin={0.75cm},
	numbers=left,
	stepnumber=1,
	firstnumber=1,
	numberfirstline=true,	
	identifierstyle=\color{black},
	keywordstyle=\color{blue}\bfseries,
	ndkeywordstyle=\color{editorGreen}\bfseries,
	stringstyle=\color{editorOcher}\ttfamily,
	commentstyle=\color{brown}\ttfamily,
	language=HTML5,
	alsolanguage=JavaScript,
	alsodigit={.:;},	
	tabsize=2,
	showtabs=false,
	showspaces=false,
	showstringspaces=false,
	extendedchars=true,
	breaklines=true,
	literate=%
	{Ö}{{\"O}}1
	{Ä}{{\"A}}1
	{Ü}{{\"U}}1
	{ß}{{\ss}}1
	{ü}{{\"u}}1
	{ä}{{\"a}}1
	{ö}{{\"o}}1
}
\renewcommand{\paragraph}{\vspace{3pt}\noindent\textbf}
\begin{document}
	
\title{\COSI (COSI) Attacks:\\ Leaking Web Site States through \xsleaks}

\author{\IEEEauthorblockN{Avinash Sudhodanan}
\IEEEauthorblockA{IMDEA Software Institute\\
avinash.sudhodanan@imdea.org}
\and
\IEEEauthorblockN{Soheil Khodayari}
\IEEEauthorblockA{CISPA Helmholtz Center for Information Security\\
soheil.khodayari@cispa.saarland}
\and
\IEEEauthorblockN{Juan Caballero}
\IEEEauthorblockA{IMDEA Software Institute\\
juan.caballero@imdea.org}}

\maketitle

\begin{abstract}
	In a \COSI (COSI) attack, an attacker convinces a victim 
into visiting an attack web page, which leverages the 
cross-origin interaction features of the victim's web browser 
to infer the victim's state at a target web site. 
Multiple instances of COSI attacks have been found in the past
under different names such as login detection or access detection attacks.
But, those attacks only consider two states (e.g., logged in or not) and 
focus on a specific browser leak method (or \xsleak).

This work shows that mounting more complex COSI attacks such as 
deanonymizing the owner of an account,
determining if the victim owns sensitive content, and
determining the victim's account type 
often requires considering more than two states. 
Furthermore, robust attacks require supporting a variety of browsers 
since the victim's browser cannot be predicted apriori.
To address these issues, 
we present a novel approach to identify and build complex
COSI attacks that differentiate more than two states and
support multiple browsers by combining multiple attack vectors, 
possibly using different \xsleaks.
To enable our approach, we introduce the concept of a COSI attack class. 
We propose two novel techniques to generalize existing COSI attack 
instances into COSI attack classes and to discover new COSI attack classes. 
We systematically apply our techniques to existing attacks, 
identifying \numAttackClasses COSI attack classes. 
As part of this process, 
we discover a novel \xsleak based on {\it window.postMessage}.
We implement our approach into \tool,
a tool to find COSI attacks in a target web site.
We apply \tool to test \nApps stand-alone web applications
and \nSites popular web sites, 
finding COSI attacks against each of them.

\end{abstract}

\section{Introduction}
\label{sec:intro}

In a \COSI (COSI) attack, the attacker's goal is to determine 
the state of a {\it victim} visiting an {\it attack page}
(e.g., \url{attack.com/index.html}),
in a {\it target web site} 
not controlled by the attacker
(e.g., \url{linkedin.com}).
The state of the victim in a target web site 
is defined, among others, 
by login status, account, and content properties.
Determining the victim's state
can have important security implications. 
For example, determining that a victim is logged into a target web site 
implies that the victim
owns an account in that site.
This is problematic for privacy-sensitive web sites such as those related
to post-marital affairs and pornography.
Determining content ownership can be used to establish if a program
committee member is reviewing a specific paper in a conference management 
system, or if the victim has uploaded some copyrighted content to an 
anonymous file sharing site. 
Determining if the victim owns a specific account, 
i.e., deanonymizing the account owner, 
enables identifying which company employee runs an anonymous blog 
criticizing the company's management.
Such state inferences are even more critical 
when the attacker is a nation state that performs censorship and 
can determine if the victim
has an account in, or is the administrator of, some prohibited web site.
The problem is aggravated by COSI attacks being web attacks,
which can be performed even when the victim employs anonymization tools
such as a virtual private network. 

In a COSI attack, the attacker convinces the victim to visit an attack page. 
The attack page includes at least one {\it {\sdURL}} (\sURL) 
from the target web site, whose response depends on the state of the visitor. 
For example, a \sURL may point to some content in
the target web site only accessible when the victim has a specific state 
such as being authenticated. 
The inclusion forces the victim's browser to send a cross-origin request
to the target web site. 
Since the request is cross-origin, 
the same-origin policy (SOP) prevents the attack page from directly 
reading the response. 
However, the attacker can leverage a {\it browser leak method} 
(also known as {\it XS-Leak}) to infer, from the cross-origin response, 
the victim's state at the target web site.

Multiple instances of COSI attacks have been found 
in the last 13 years by both security analysts
(e.g.,~\cite{GrossmanCosiImage2006,Grossmancosijserror2006,Evanscsslo2008,Evanssearchtiming2009,HomakovCSPbug2013,Linuscoldsocial2016}) 
and academics
(e.g.,~\cite{BortzTiming2007WWW,VanGoethemClock2015CCS,GulyasCSPLO2018WPESextend,GelernterCSSA2015CCS,SanchezBakingTimer2019ACSAC}),
with roughly half of them being presented in the last four years, 
and several in 2019 
(e.g.,~\cite{StaicuLeakyImg2019SEC,MasasImpervaWLStat2019,SanchezBakingTimer2019ACSAC}). 
However, they have previously been considered as sparse attacks
under different names such as
login detection attacks~\cite{Linuscoldsocial2016,Grossmancold2008whoseproblem,Grossmancoldmany2012,SanchezBakingTimer2019ACSAC}, 
login oracle attacks~\cite{JorgSOPEval2017SEC,LekiesXSSI2015SEC}, 
cross-site search attacks~\cite{GelernterCSSA2015CCS}, 
URL status identification attacks~\cite{LeeAppCacheErr2015NDSS},
and cross-site frame leakage attacks~\cite{MasasImpervaWLStat2019}. 
As far as we know, we are the first to systematically study these 
attacks and group them under the same COSI attack denomination. 

Previous works have several limitations. 
First, they consider two states.
For example, login detection attacks differentiate if the victim is
logged in or not, and access detection attacks if the victim 
has previously accessed a site or not. 
However, sites typically have more than two states. 
Considering only two states limits the type of attacks that can be launched, 
and can introduce false positives, 
e.g., determining that a victim is logged in when he is not.
A second limitation is that they often test attacks 
only on one browser, thus the attack may not work on other browsers. 
To address both issues, we present a novel approach 
to identify and build complex COSI attacks by combining multiple attack vectors 
in order to handle more than two states and multiple browsers.  
For example, our approach identifies a COSI attack against HotCRP that 
determines if the victim, 
i.e., a program committee member using Chrome, Firefox, or Edge 
is the reviewer of a submitted paper. 
This attack involves multiple states 
(e.g., author, reviewer, logged out) and requires two COSI attack vectors: 
one to determine if the victim is logged in and another to determine 
if a logged victim is reviewing the paper.

A third limitation is that they focus on a specific \xsleak.
Instead, our approach is generic; 
it supports all known \xsleaks and can easily accommodate new ones. 
For example, it incorporates a novel \xsleak we have discovered 
based on {\it window.postMessage},
which affects popular sites such as 
\url{blogger.com}, \url{ebay.com}, \url{reddit.com}, and \url{youtube.com}.
At the core of our generic approach is 
the concept of a {\it COSI attack class}, 
which defines the \sURLs that can be attacked using a specific 
\xsleak, the affected browsers, and the set of {\it inclusion methods} 
(i.e., HTML tags and DOM methods) that can be used to include the \sURL 
in the attack page. 
To identify attack classes we propose a novel generalization technique 
that given a previously known COSI attack,
generalizes it into an attack class that covers many other attack 
variants.
We also propose an amplification technique that identifies 
previously unknown variations, 
e.g., attack classes using different inclusion methods. 
We systematically explore the literature to identify 
previously known COSI attack instances and
apply our generalization and amplification techniques on them.
This process identifies \numAttackClasses COSI attack classes,
of which \numKnowAttackClasses generalize prior attacks and 
\numNewAttackClasses are new variations.

We implement our approach into a tool called \tool, 
publicly available as part of the open-source 
ElasTest platform~\cite{elastest}. 
Given as input a target web site and state scripts defining the user states
at the target web site, \tool identifies \sURLs in the target web site, 
tests if those \sURLs can be attacked using any of the \attackClasses, and
produces attack pages that combine multiple attack vectors 
to uniquely identify a state.
We have applied \tool to \nTargets targets: 
\nApps stand-alone web applications 
(HotCRP, GitLab, GitHub, and OpenCart) and 
\nSites popular web sites.
\tool discovers at least one COSI attack against all of them;
it finds login detection attacks against all \nTargets targets, 
account deanonymization attacks in \nUserIdentAtks, 
account type detection attacks in \nAccTypeAtks,
SSO status attacks in \nSSOAtks, 
and access detection attacks in \nAccsDetAtks.
The attacks include, among others, 
deanonymization attacks for determining if the victim is 
the reviewer of a paper in HotCRP,
owns a blog in \url{blogger.com}, 
an account in \url{pornhub.com}, or
a GitLab/GitHub repository. 

\noindent The following are the main contributions of this paper:

\begin{itemize}
    \item We present a novel approach to identify and build complex 
        COSI attacks that differentiate more than two states and 
        support multiple browsers.
        To enable our approach we propose COSI attack classes, 
        which define the SD-URLs and browsers that can be
        attacked using an XS-Leak and a set of inclusion methods.

    \item We discover a novel \xsleak based on 
        {\it window.postMessage} that affects the three major browsers and can 
        be leveraged to attack popular web sites.

    \item We propose two techniques to generalize known COSI attack instances 
        into COSI attack classes and to discover new variations. 
        We perform the first systematic study of COSI attacks and apply 
        our techniques to them, identifying \attackClasses, 
        of which \numKnowAttackClasses generalize prior attacks and
        \numNewAttackClasses are new variations.

  \item We implement our approach into \tool, 
        a tool to find COSI attacks in a target web site. 
        We apply \tool to \nTargets targets including 
        stand-alone web applications and popular live sites.
        We find COSI attacks against all of them, 
        enabling account deanonymization, account type inference, 
        SSO status, login detection, and access detection.

  \item We have released \tool as part of the security service 
        of the ElasTest open-source platform 
        for testing cloud applications~\cite{elastest}.

\end{itemize}
\begin{table}[t]
  \caption{Examples of user states in a target web site.}
  \label{table:states}
  \centering
  \scriptsize
  \begin{tabular}{ll}\toprule
    \textbf{State Attribute} & \textbf{Possible Values} \\
    \hline
    Login Status & (a) Logged in \\
                 & (b) Not logged in \\ 
    \hline
    Single Sign-On Status & (a) Logs in via a specific SSO service \\
                          & (b) Logs in via another SSO service \\ 
    \hline
    Access Status & (a) Has previously accessed \\ 
                  & (b) Has not previously accessed \\
    \hline
    Account Type & (a) Has a premium account \\
                 & (b) Has a regular account \\ 
    \hline
    Account Age Category & (a) Age above a certain threshold \\
                         & (b) Age below a certain threshold \\ 
    \hline
    Account Ownership & (a) Owner of a specific account \\
                      & (b) Not the owner of an account \\
    \hline
    Content Ownership & (a) Owner of a specific content \\
                      & (b) Not the owner of a content \\
    \hline
  \end{tabular}
\end{table}

\section{Overview}
\label{sec:overview}

This Section provides an overview of COSI attacks. 
Section~\ref{sec:state} details the user state at a target web site.
Section~\ref{sec:phases} describes the two phases of a COSI attack. 
Section~\ref{sec:multistates} discusses handling more than two states. 
Finally, Section~\ref{sec:threat} presents the COSI attack threat model.

\subsection{User State}
\label{sec:state}

Most web sites have accounts owned by a user and identified by a username.
In this paper a user is a person who visits a target web site and
may or may not own an account in that site;
it should not be confused with a username that identifies an account.
Accounts are often anonymous,
i.e., the person that owns the account is unknown.
Deanonymizing an account means linking its username to the
person owning the account.
Web sites that do not have accounts often define sessions to identify
users that visit them repeatedly.
In those sites a session acts as an account for our purposes.

In a COSI attack, the attacker's goal is to infer the state of a 
victim user with respect to a target web site,
not controlled by the attacker.
The state of a user at a target web site is defined by 
the values of status, account, and ownership state attributes. 
Example state attributes are provided in Table~\ref{table:states}.
The values of those state attributes define, at a given time, 
what content the user can access (or receives) from the target site.
Status attributes include 
whether the user is logged in, logged out, 
logged in using a specific single sign-on (SSO) service, or
has an ongoing session 
(i.e., in sites without user accounts).
Account attributes include 
the account type 
(e.g., regular, premium, administrator), 
the account age category
(e.g., under-age user with restricted access).
Ownership attributes include whether the user is the owner of some 
specific account and 
whether he owns some content stored in the site
(e.g., a PDF paper in a conference management system).

The attributes that define the user's state are specific to each target site. 
Any of those attributes may be targeted by an attacker with 
different, often critical, security implications.
For example, COSI attacks targeting the login status can be used 
by an oppressive regime to determine if the victim is logged in 
(and thus owns an account) in a censored site~\cite{CardwellColdScode2011}, 
despite the victim using a VPN.
They can also be used to blackmail users owning
accounts in privacy-sensitive sites such as those related to 
pornography~\cite{CrawleyPorn2018blackmail} and 
post-marital affairs~\cite{HernAshleymadd2016blackmail}.
Furthermore, they may be used as an initial step for 
Cross-Site Request Forgery (CSRF)~\cite{BarthCSRF2008CCS}
or Cross-Site Scripting (XSS)~\cite{LekiesDOMXSS2013CCS} attacks.
Attacks on access status have similar implications than those on 
login status for sites without user accounts.
For example, they could be used to determine if a user previously 
visited a forbidden site \cite{SanchezBakingTimer2019ACSAC}. 

COSI attacks targeting ownership are highly impactful.
Content ownership can be used to determine 
if a program committee member is reviewing a specific paper, or if a user has 
uploaded some copyrighted content to an anonymous file sharing site.
Account ownership can be used for deanonymizing the account in a closed-world setting, 
i.e., determining which of $n$ known persons owns a specific account. 
Such closed-world deanonymization can be used to determine 
which company employee is the owner of an anonymous blog 
highly critical with the company's management.  

Attacks that target account type, account age category, and login status
can be used to fingerprint the 
victim~\cite{Homakovcsp4evil2014,Cspfingerprinting2018}, 
and applied for targeted advertising by a malicious publisher 
in an open-world setting (where the set of users is unknown).
Finally, knowledge of the SSO service used by the victim can be used to 
exploit a vulnerability in that 
SSO~\cite{ArmandoSAML2008FMSE,BhargavanSSO2012CSF,Wang2012SnP}.

\paragraph{State scripts.}
In this work, we capture states at a target site using 
state scripts that can be executed to automatically log into the 
target site using a configurable browser and 
the credentials of an account with a specific configuration.
For example, we may create multiple user accounts 
with different configurations, 
e.g., premium and free accounts, two users that own different blogs, or 
authors that have submitted different papers to a conference management system.
We also create a state script for the logged out state. 

\subsection{COSI Attack Overview}
\label{sec:phases}

In a COSI attack, the attacker convinces a victim to visit an attack page. 
The attack page leverages the cross-origin functionalities of the 
victim's web browser to infer the victim's state at a target web site.
A COSI attack comprises of two phases: {\it preparation} and {\it attack}.
 
\paragraph{Preparation.}
The goal of the preparation phase is to create an 
attack page that when visited by a victim will leak the victim's state 
at the target web site.
An attack page implements at least one, possibly more, attack vectors. 
Each attack vector is a triplet of 
a {\it \sdURL} from the target web site, 
an {\it inclusion method} to embed the \sURL in the attack page, and
an {\it attack class} that defines, among others,  
a {\it leak method} (or {\it \xsleak}) that interacts with the victim's browser to 
disclose a victim's state at the target site.
An attack page may contain multiple attack vectors. 
For example, it may need to chain attack vectors to uniquely 
distinguish a state, 
e.g., one to identify if the victim is logged in, and another to identify 
if a logged victim has a premium account.

We say that a URL is state-dependent if, when requested through HTTP(S), 
it returns different responses depending on the state it is visited from.
Note that it is not needed that each state returns a different response. 
For example, if there are 6 states and two different responses, 
each for three states, the URL is still state-dependent.
The \sURL is included by the attack page using an inclusion method 
such as an HTML tag (e.g., {\it img}, {\it script}) or 
a browser DOM method (e.g., {\it window.open}). 
When the attack page is visited by the victim, 
the inclusion method forces the victim's browser to automatically 
request the \sURL from the target site.
The specific response received depends on the victim's current state. 
\sURLs are very common in web applications. 
For example, in many web applications, 
sending a request for a profile's picture
will return an image if the user is logged in, 
and an error page, or a redirection to the login page, otherwise.
Similarly, in a blog application, a new post can only be added if the user 
is both logged in and the owner of the blog.

The request induced by the attack page for a 
{\sURL} at the target site is cross-origin, 
and thus controlled by the  Same-Origin Policy (SOP)~\cite{Zalewskisop2008}.
The SOP prevents the attack page from directly reading the 
contents of a cross-origin response~\cite{Barth2011web}. 
However, there exist \xsleaks that allow bypassing a browser's SOP to disclose information about cross-origin responses.
For example, the {\it EventsFired} \xsleak distinguishes responses 
to \sURLs that trigger a callback in one state (e.g., {\it onload}) and 
another callback (e.g., {\it onerror}), or no callback, in another state \cite{GrossmanCosiImage2006}.

While a target site may contain many {\sURL}s, 
only a subset of those may be useful to mount a COSI attack. 
One main challenge with \xsleaks is that their behavior may depend 
on the target browser and the inclusion method used.
Unfortunately, this key concept is missing from prior works 
presenting COSI attack instances.
In this work, we introduce the concept of a 
{\it COSI attack class}, which defines 
the two different responses to a \sURL that can be distinguished 
using a \xsleak, 
the possible inclusion methods that can be used in conjunction 
with the \xsleak, and the browsers affected.
Attacks classes are independent of the target site states and 
thus can be used to mount attacks against different targets. 
Section~\ref{sec:classes} describes our approach to identify attack 
classes and the \numAttackClasses COSI attack classes we identified.

Based on the attack classes, we propose a novel approach to 
detect COSI attacks.
Our approach first collects the responses to the same URL 
from different states. 
\sURLs will be the ones that produce different responses in some states.
Each pair of different responses coming from distinct states 
is matched with the list of known attack classes. 
If a matching attack class is found, then an attack vector can be 
built to distinguish the responses (and thus the states that produce them) 
that uses that \sURL, the \xsleak in the attack class, and one 
of the inclusion methods defined by the attack class.
Since there may be $n > 2$ states that need to be distinguished, 
the process repeats until sufficient attack vectors are identified 
to uniquely distinguish the target state to be attacked. 
We have implemented this approach into \tool, 
a tool to detect COSI attacks, detailed in Section~\ref{sec:tool}.

\paragraph{Attack.} 
In the attack phase, the attacker convinces the victim 
into visiting the attack page. 
This can be achieved in multiple ways. 
One possibility is sending an email with the attack page URL and 
text to convince the victim to click on it. 
Such targeted attack requires the victim's email, 
but allows identifying the state of a specific person, 
e.g., deanonymizing the owner of an account.
Another possibility is a watering-hole approach where the attacker injects 
the attack page URL into a vulnerable page that victims are likely to visit.
Such attack allows identifying the state of a visitor, 
but does not identify who the visitor is.
The method used to convince the victim to visit 
the attack page is outside the scope of this paper.
When the attack page is loaded at the victim's browser, 
it checks the browser used by the victim, 
delivers suitable attack vectors, and
reports back the leaked victim's state.

\subsection{Beyond Two States}
\label{sec:multistates}

Current COSI attacks targeting login or access detection consider 
only two states.
However, most web sites have more than two states, 
e.g., logged in users with different permissions. 
Considering only two states introduces some issues. 
First, it limits the type of attacks, 
preventing attacks that target finer-grained states 
such as account type or content ownership.
Furthermore, it can introduce false positives, 
which is best illustrated with an example.

In 2015, Lee et al.~\cite{LeeAppCacheErr2015NDSS} presented a novel 
AppCache \xsleak (described in Section~\ref{sec:classes}) 
that enabled login detection.
One of their login detection attacks targeted the NDSS 2015 HotCRP installation.
The \sURL \url{https://ndss2015.ccs.neu.edu/paper/<paper-no>} returned 
a success HTTP status code when the victim was logged into HotCRP and 
an error status code otherwise. 
That difference could be identified using the AppCache \xsleak.
In reality, the HotCRP access control is more fine-grained
and the information of a paper can only be accessed by its 
authors or by reviewers, 
but not by other authors who would also receive an error. 
Thus, their attack could incorrectly identify an authenticated victim, 
who happened to be an author of another paper, as not being authenticated.
Such false positives could be avoided if they could guarantee 
that victims would not be authors 
(e.g., not sending authors an email with the attack page URL),
but authors are only known to the conference administrators.

\begin{lstlisting}[float,style=htmlcssjs,caption={Running example attack page for deanonymizing the reviewer of a paper in HotCRP.},label={lst:attackpage}]
<!DOCTYPE html><html>
//Launch attack when page loads
<body onload="attack()"><script>
//SD-URLs used in the attack vectors
site = "https://conf.hotcrp.com"
loginURL = site+"/offline.php?downloadForm=123";
reviewURL = site+"/api.php/review?p=123";
//Object for storing fired events
evnts = {"obj": [], "lnk" : [], "embd" : []}
function attack() {
  // Login detection on all browsers
	EF_XctoObject();
  // Reviewer deanonymization
	if (detectBrowser() == "Chrome") {
		EF_StatusErrorLink();
	}
	else { EF_StatusErrorObject(); }
	sendToAttkr(evnts); //send events to attacker
}
function EF_XctoObject() {
	tag = document.createElement("object");
	tag.setAttribute("data", loginURL);
	tag.setAttribute("rel", "stylesheet");
	tag.onload = function(){
		evnts["obj"].push("onload");
	}
	document.body.appendChild(tag);
}
function EF_StatusErrorLink(){...}
function EF_StatusErrorObject(){...}
</script></body></html>
\end{lstlisting}

\paragraph{Running example.}
As running example we use a reviewer deanonymization attack \tool 
found on HotCRP, which was acknowledged and fixed. 
Listing~\ref{lst:attackpage} shows a simplified version of the attack page 
produced by \tool that we sent to HotCRP developers to report the attack. 
It identifies if the visiting victim 
is the reviewer of paper \#123
submitted to \url{https://conf.hotcrp.com}.
Since HotCRP has multiple states 
(e.g., logged in, author, reviewer, reviewer of a specific paper) 
and we want to support the major browsers (Chrome, Firefox, Edge),
the attack page requires three attack vectors
executed when the attack page is loaded (Line 3).
It first runs an attack vector for determining the 
victim's login status, which works regardless if the victim's browser is 
Chrome, Firefox, or Edge (Lines 12, 20-28). 
This attack vector includes \sURL 
\url{https://conf.hotcrp.com/offline.php?downloadForm=123} 
with the \texttt{object} HTML tag and uses the \textit{EventsFired} \xsleak: 
if the victim is logged into the site, no events are triggered,
otherwise the \textit{onload} event is triggered.
Then, it executes the attack vectors for 
reviewer deanonymination, which differ 
for Chrome (Line 15) and Firefox/Edge (Line 17).
These attack vectors are not detailed for brevity, 
but both use the \textit{EventsFired} \xsleak with different inclusion 
methods for the same \sURL \url{https://conf.hotcrp.com/api.php/review?p=123}, 
which returns a success HTTP status code if the victim has submitted a review 
for paper \#123, and an error HTTP status code otherwise.
\subsection{Threat Model}
\label{sec:threat}

This section describes the COSI attack threat model, 
detailing the assumptions we make about each actor. 

\paragraph{Attacker.}
We assume that the attacker can trick victims into loading the attack page 
on their web browsers.
During preparation, the attacker has the ability to 
create and manage different accounts at the target web site, 
or in a local installation of the target's web application. 
The attacker controls an attack web site where he
can add arbitrary pages.
Finally, we assume the attacker can identify the victim's browser version 
(e.g., from the User-Agent header) to select the right attack vector.

\paragraph{Victim.}
The victim uses a fully up-to-date web browser and 
can be lured by the attacker into visiting the attack webpage. 
We assume that the victim logs into the target web site 
with the same web browser used to visit the attack page.

\paragraph{Target site.}
The target site contains at least one \sURL for which the attacker knows 
an attack class.
The target site does not suffer from any known vulnerabilities.
In particular, resources containing sensitive information 
are protected from direct cross-origin reads, 
i.e., the target site does not contain CORS 
misconfigurations~\cite{LekiesCORS2011W2SP}, 
cross-site scripting \cite{LekiesDOMXSS2013CCS}, or 
\xssi vulnerabilities \cite{LekiesXSSI2015SEC}.
\section{COSI Attack Classes}
\label{sec:classes}

A key concept in our approach are COSI attack classes.
A COSI attack class is a 6-tuple that comprises of a
class name, signatures for two groups of responses that can be 
distinguished using the attack class, 
an \xsleak, 
a list of inclusion methods that can be used to embed the \sURL 
in an attack page, and the list of affected browsers.
It captures which \sURLs can be used for
building an attack vector against the affected browsers
using the \xsleak and one of the inclusion methods defined.
A reader could think that an attack class should simply correspond to an 
\xsleak. 
However, the behavior of some \xsleaks depends on the target browser and 
the inclusion method used. 
Depending on those two parameters, the set of affected \sURLs differs. 
Thus, identifying attack classes is fundamental for determining 
whether and how a given \sURL can be attacked.  
This section first presents our approach to discover COSI attack classes
in Section~\ref{sec:classDiscovery} and then details the \attackClasses 
identified in Section~\ref{sec:classDescription}.

\subsection{Discovering Attack Classes}
\label{sec:classDiscovery}

Our process to discover COSI attack classes comprises of three main steps: 
(1) identify and validate previously proposed COSI attack instances; 
(2) generalize known COSI attack instances into COSI attack classes; and
(3) discover previously unknown attack classes. 

\paragraph{Identifying attack instances.}
We have performed a systematic survey of COSI attack instances 
presented in prior work under different names. 
This process identified \numPriorWorks prior works,
listed in Table~\ref{tab:attacks} and described in Section~\ref{sec:related}. 
Out of those, 11 are blog posts, 10 are academic papers, one is a bug report, 
and the last one is a project simultaneous to our work 
that tries to enumerate all known
\xsleaks~\cite{XsleaksCollection2019}.
Those \priorWorks presented \attackInstances. 
All attack instances could be validated in at least one recent browser version.
To validate an attack instance we manually create a test attack page 
based on the available information. 
The test attack page includes a URL from a test application 
we have designed to return custom responses to an incoming request. 
Requests to the test application define how the response should look 
(i.e., which headers and body to return). 
In this step, we configured our test application to return the 
responses described in the work presenting the attack.
This enables validating attack instances even when the \sURL 
used in the attack was no longer active.

\paragraph{Generalizing instances into classes.}
Generalizing a COSI attack instance into a COSI attack class comprises of 
two steps. 
First, identifying the set of responses to the inclusion method 
used in the attack instance, 
that still trigger the same observable difference in the browser 
(e.g., onload/onerror or different object property values). 
Then, checking if the observable difference still manifests with 
other inclusion methods and browsers.
The generalization uses the test application
to control the response received from a potential target site.
We illustrate it using an attack instance of the 
\textit{EF-StatusErrorObject} attack class.
The generalization starts with the response that triggers the onload callback 
and tries to modify each response element (header or body) to a different value.
If the modification still triggers the onload callback, then the element can 
be ignored.
In our example, all fields can be ignored, except the status code that it 
should be 200 and the content-type that should not correspond to 
an audio or video. 
The generalization then repeats for the response that triggers the onerror
callback, returning that the status code should not be success (200) or 
redirection (3xx), but other values for the status code, headers, and body 
do not matter. 
Once the responses are generalized, it tests whether other inclusions methods
still trigger the same observable difference.
For this, it tests the \textit{window.open()} method and 
the 13 HTML tags that enable resource inclusion without user intervention, 
shown in Table~\ref{tab:tags}.
Finally, it checks if the leak manifests in other browsers.
Table~\ref{tab:attacks} shows that the \attackInstances examined 
belonged to 15 attack classes, i.e., many were duplicates. 

\begin{table}[t]
  \small
  \caption{HTML tags supporting resource inclusion.}
  \label{tab:tags}
  \scriptsize
  \centering
  \begin{tabular}{lll}\toprule
    \textbf{Tag} & \textbf{Attribute} & \textbf{Included Resource's Type}\\ \midrule
    \texttt{applet} & \texttt{code} & \multicolumn{1}{l}{Applet}\\
    \texttt{audio} & \texttt{src} & \multicolumn{1}{l}{Audio}\\
    \texttt{embed} & \texttt{src} & \multicolumn{1}{l}{Defined in \texttt{type} attribute}\\
    \texttt{frame} & \texttt{src} & \multicolumn{1}{l}{Typically web pages}\\
    \texttt{iframe} & \texttt{src} & \multicolumn{1}{l}{Typically web pages}\\
    \texttt{img} & \texttt{src} & \multicolumn{1}{l}{Image} \\
    \texttt{input} & \texttt{src} & \multicolumn{1}{l}{Image (when attr. \texttt{type} = ``picture")}\\
    \texttt{link} & \texttt{href} & \multicolumn{1}{l}{Defined in \texttt{rel} and \texttt{type} attributes}\\
    \texttt{object} & \texttt{data} & \multicolumn{1}{l}{Defined in \texttt{type} attribute}\\
    \texttt{script} & \texttt{src} & \multicolumn{1}{l}{JS}\\
    \texttt{source} & \texttt{src} & \multicolumn{1}{l}{Audio/Video}\\
    \texttt{track} & \texttt{src} & \multicolumn{1}{l}{WebVTT \cite{webvtt}}\\
    \texttt{video} & \texttt{poster} & \multicolumn{1}{l}{Image}\\
    \texttt{video} & \texttt{src} & \multicolumn{1}{l}{Video}\\
    \bottomrule
  \end{tabular}
\end{table}

\paragraph{Discovering new attack classes.}
The test application allows systematically exploring 
combinations of header and body values in responses. 
For each response, browser events and DOM values are logged. 
Pairs of responses that produce observable differences 
(e.g., trigger different callbacks),
and do not match existing attack classes, correspond to new attack instances, 
and are generalized as above.
Overall, we discovered \numNewAttackClasses new attack classes, 
of which 12 use the EventsFired (i.e., onload/onerror) \xsleak, 
8 use the Object Property \xsleak, and 
1 uses a completely novel \xsleak based on postMessage.
\begin{table*}
	\caption{COSI attack classes.}
	\label{tab:classes}
	\centering
	\tiny
	\renewcommand{\arraystretch}{1.5}
	\begin{tabular}{llllllll}\toprule
		\multicolumn{1}{l}{\textbf{Class}} & \multicolumn{2}{c}{\textbf{\sURL Responses}} & \multicolumn{2}{c}{\textbf{Attack Page's Logic}} & \multicolumn{3}{c}{\textbf{Browsers}} \\

		& \textit{Response A}& \textit{Response B}& \textit{Inclusion Methods} & \textit{Leak Method} & \multicolumn{1}{c}{\textit{Firefox}} & \multicolumn{1}{c}{\textit{Chrome}} & \multicolumn{1}{c}{\textit{Edge}}\\ \midrule
		
		EF-StatusErrorScript & \multicolumn{1}{p{3cm}}{sc = 200, ct = text/javascript} & \multicolumn{1}{p{3cm}}{sc = (4xx OR 5xx)} & \texttt{script src=URL} & \multicolumn{1}{l}{[onload] / [onerror]} & \Y & \Y & \Y \\
		
		EF-StatusErrorObject & \multicolumn{1}{p{3cm}}{sc = 200, ct $\ne$ (audio OR video)} & \multicolumn{1}{p{3cm}}{sc $\ne$ (200 OR 3xx)} & \texttt{object data=URL} & \multicolumn{1}{l}{[onload] / [onerror]} & \Y & \N & \N \\

		EF-StatusErrorEmbed & \multicolumn{1}{p{3cm}}{sc = 401, ct = (text/html)} & \multicolumn{1}{p{3cm}}{sc $\ne$ 401, ct = (text/html)} & \texttt{embed src=URL} & \multicolumn{1}{l}{[] / [onload]} & \N & \N & \Y \\ %
					
		EF-StatusErrorLink & \multicolumn{1}{p{3cm}}{sc = (200 OR 3xx), ct $\ne$ text/html} & \multicolumn{1}{p{3cm}}{sc $\ne$ (200 OR 3xx)} & \multicolumn{1}{p{3cm}}{\texttt{link href=URL rel=prefetch}} & \multicolumn{1}{l}{[onload] / [onerror]} & \N & \Y & \N \\
		
		EF-StatusErrorLinkCss & \multicolumn{1}{p{3cm}}{sc = (200 OR 3xx), ct = text/css} & \multicolumn{1}{p{3cm}}{sc $\ne$ (200 OR 3xx), ct $\ne$ text/css} & \multicolumn{1}{p{3cm}}{\texttt{link href=URL rel=stylesheet}} & \multicolumn{1}{l}{[onload] / [onerror]} & \Y & \Y & \N \\
		
		EF-RedirStatLink & \multicolumn{1}{p{3cm}}{sc = 3xx} & \multicolumn{1}{p{3cm}}{sc $\ne$ 3xx, cto = nosniff, ct $\ne$ (text/css OR text/html)} & \multicolumn{1}{p{3cm}}{\texttt{link href=URL rel=stylesheet}} & \multicolumn{1}{l}{[onload] / [onerror]} & \N & \Y & \N \\
		
		EF-StatusErrorIFrame & \multicolumn{1}{p{3cm}}{sc = (200 OR 3xx OR 4xx or 5xx), ct= (text/javascript OR text/css)} & \multicolumn{1}{p{3cm}}{sc = (200 OR 3xx OR 4xx or 5xx), ct $\ne$ (text/javascript OR text/css)} & \texttt{iframe src=URL} & \multicolumn{1}{l}{[] / [onload]} & \N & \N & \Y \\

		EF-NonStdStatusErrorIFrame & \multicolumn{1}{p{3cm}}{sc = (200 OR 3xx OR 4xx or 5xx), ct = (text/javascript OR text/css)} & \multicolumn{1}{p{3cm}}{sc = 999} & \texttt{iframe src=URL} & \multicolumn{1}{l}{[] / [onload]} & \N  & \N & \Y \\
		
		EF-CDispIFrame & \multicolumn{1}{p{3cm}}{sc = 200, cd = attachment} & \multicolumn{1}{p{3cm}}{cd $\ne$ attachment} & \texttt{iframe src=URL} & \multicolumn{1}{l}{[] / [onload]} & \N & \Y & \N \\
		
		EF-CDispStatErrIFrame & \multicolumn{1}{p{3cm}}{sc = (4xx OR 5xx), cd = attachment} & \multicolumn{1}{p{3cm}}{sc = (4xx OR 5xx), cd $\ne$ attachment} & \texttt{iframe src=URL} & \multicolumn{1}{l}{[] / [onload]} & \Y & \N & \N \\
		
		EF-CDispAthmntIFrame & \multicolumn{1}{p{3cm}}{sc = 200, cd = attachment} & \multicolumn{1}{p{3cm}}{$\neg$(sc = 200, cd = attachment)} & \texttt{iframe src=URL} & \multicolumn{1}{l}{[] / [onload]} & \N & \Y & \N \\
		
		EF-XctoScript & \multicolumn{1}{p{3cm}}{sc = 200, xcto disabled, ct = (text/html OR text/css OR application/pdf)} & \multicolumn{1}{p{3cm}}{sc = 200, xcto = nosniff, ct = (text/html OR text/css OR application/pdf)} & \texttt{script src=URL} & \multicolumn{1}{l}{[onload] / [onerror]} & \Y & \N & \Y \\
		
		EF-XctoObject & \multicolumn{1}{p{3cm}}{sc = 200, xcto disabled, ct = (text/html OR text/css OR application/json)} & \multicolumn{1}{p{3cm}}{sc = 200, xcto = nosniff, ct = (text/html OR text/css OR application/json)} & \texttt{object data=URL} & \multicolumn{1}{l}{[onload] / [ ]} & \Y & \Y & \Y \\ 
				
		EF-CtMismatchObject & \multicolumn{1}{p{3cm}}{sc = 200, ct = X} & \multicolumn{1}{p{3cm}}{sc = 200, ct = Y} & \multicolumn{1}{p{3cm}}{\texttt{object data=URL typesmustmatch type=X}} & \multicolumn{1}{l}{[onload] / [onerror]} & \Y & \N & \N \\
		
		EF-CtMismatchScript & \multicolumn{1}{p{3cm}}{sc = 200, ct = (text/javascript)} & \multicolumn{1}{p{3cm}}{sc = 200, xcto = nosniff, ct $\ne$ (text/javascript)} & \texttt{script src=URL} & \multicolumn{1}{l}{[onload] / [onerror]} & \Y & \N & \Y \\
		
		EF-CtMismatchImg & \multicolumn{1}{p{3cm}}{sc = (200 OR 3xx OR 4xx OR 5xx), ct = image} & \multicolumn{1}{p{3cm}}{sc = (200 OR 3xx OR 4xx OR 5xx), ct $\ne$ image} & \texttt{img src=URL } & \multicolumn{1}{l}{[onload] / [onerror]} & \N & \Y & \Y\\

		EF-CtMismatchAudio & \multicolumn{1}{p{3cm}}{sc = (200 OR 3xx OR 4xx OR 5xx), ct = audio} & \multicolumn{1}{p{3cm}}{sc = (200 OR 3xx OR 4xx OR 5xx), ct $\ne$ audio} & \texttt{audio src=URL } & \multicolumn{1}{p{1.4cm}}{$\neg$[onerror OR onsuspend] / [onerror OR onsuspend]} & \N & \Y & \N\\
		
		EF-CtMismatchVideo & \multicolumn{1}{p{3cm}}{sc = (200 OR 3xx OR 4xx OR 5xx), ct = video} & \multicolumn{1}{p{3cm}}{sc = (200 OR 3xx OR 4xx OR 5xx), ct $\ne$ video} & \texttt{video src=URL } & \multicolumn{1}{p{1.4cm}}{$\neg$[onerror OR onsuspend] / [onerror OR onsuspend]} & \Y & \N & \N\\
				
		EF-XfoObject & \multicolumn{1}{p{3cm}}{sc = 200, xcto = text/*, xfo is disabled} & \multicolumn{1}{p{3cm}}{sc = 200, xfo is enabled} & \texttt{object data=URL} & \multicolumn{1}{l}{[] / [onload]} & \N & \Y & \N\\
		
		EF-CacheLoadCheck & \multicolumn{1}{p{3cm}}{bdy = includes URL A} & \multicolumn{1}{p{3cm}}{bdy = does not include URL A} & \multicolumn{1}{p{3cm}}{\texttt{Send error req to URL A, link rel=preload href=URL, img src=URL A, send error req to URL A}} & \multicolumn{1}{l}{[onload]/[onerror]} & \Y & \Y & \N\\
		
		OP-LinkSheet & \multicolumn{1}{p{3cm}}{sc = 200, ct = text/css, bdy = CSS-like} & \multicolumn{1}{p{3cm}}{sc = 200, ct $\ne$ text/css, bdy $\ne$ CSS-like} & \multicolumn{1}{p{3cm}}{\texttt{link rel=stylesheet href=URL}} & \multicolumn{1}{p{1.5cm}}{sheet} & \N & \N & \Y\\
		
		OP-LinkSheetStatusError & \multicolumn{1}{p{3cm}}{sc = (200 OR 3xx), ct $\ne$ text/css} & \multicolumn{1}{p{3cm}}{sc $\ne$ (200 OR 3xx)} & \multicolumn{1}{p{3cm}}{\texttt{link rel=stylesheet href=URL}} & \multicolumn{1}{p{1.5cm}}{sheet} & \N & \N & \Y\\
		
		OP-ImgDimension & \multicolumn{1}{p{3cm}}{sc = (200 OR 3xx OR 4xx OR 5xx), ct = image, bdy = image with dimension \textit{A}} & \multicolumn{1}{p{3cm}}{sc = (200 OR 3xx OR 4xx OR 5xx), ct = image, bdy = image with dimension \textit{B}} & \multicolumn{1}{p{3cm}}{\texttt{img src=URL}} & \multicolumn{1}{p{1.5cm}}{height, width, naturalHeight, naturalWidth} & \Y & \Y & \Y\\

		OP-VideoDimension & \multicolumn{1}{p{3cm}}{sc = (200 OR 3xx OR 4xx OR 5xx), bdy = video with dimension \textit{A}} & \multicolumn{1}{p{3cm}}{sc = (200 OR 3xx OR 4xx OR 5xx), body = (video with dimension \textit{B} OR body not video)} & \multicolumn{1}{p{3cm}}{\texttt{video src=URL}} & \multicolumn{1}{p{1.5cm}}{videoHeight, videoWidth} & \Y & \Y & \Y\\
		
		OP-WindowDimension & \multicolumn{1}{p{3cm}}{sc = (200 OR 3xx OR 4xx OR 5xx), bdy = PDF} & \multicolumn{1}{p{3cm}}{sc = (200 OR 3xx OR 4xx OR 5xx), body $\ne$ PDF} & \multicolumn{1}{p{1.5cm}}{\texttt{frame src=URL}} & \multicolumn{1}{p{1.5cm}}{height, width} & \N & \N & \Y\\
		
		OP-MediaDuration & \multicolumn{1}{p{3cm}}{sc = 200, ct = (audio or video), bdy = audio/video with duration \textit{A}} & \multicolumn{1}{p{3cm}}{sc = 200, ct = (audio OR video), bdy = audio/video with duration \textit{B}} & \multicolumn{1}{p{1.5cm}}{\texttt{audio/video src=URL}} & \multicolumn{1}{p{1.5cm}}{duration} & \Y & \Y & \Y\\

		OP-ImgCtMismatch & \multicolumn{1}{p{3cm}}{sc = 2xx, ct = image} & \multicolumn{1}{p{3cm}}{sc = 4xx, ct $\ne$ image} & \multicolumn{1}{p{1.5cm}}{\texttt{img src=URL}} & \multicolumn{1}{p{1.5cm}}{height, width, naturalHeight, naturalWidth} & \Y & \N & \Y\\

		OP-MediaCtMismatch & \multicolumn{1}{p{3cm}}{sc = 200, ct = (audio OR video)} & \multicolumn{1}{p{3cm}}{ ct $\ne$ (audio OR video)} & \multicolumn{1}{p{3cm}}{\texttt{audio/video src=URL}} & \multicolumn{1}{p{1.5cm}}{networkState, readyState, buffered, paused, duration, seekable} & \Y & \Y & \Y\\

		OP-FrameCount & \multicolumn{1}{p{3cm}}{sc = 200, ct = text/html, bdy = HTML with numFrames \textit{A}} & \multicolumn{1}{p{3cm}}{sc = 200, ct = text/html, xfo is disabled, bdy = HTML with numFrames \textit{B}} & \multicolumn{1}{p{3cm}}{\texttt{iframe src=URL}, (\texttt{form}, \texttt{iframe})} & \multicolumn{1}{p{1.5cm}}{contentWindow.length} & \Y & \Y & \Y\\

		OP-MediaStatus & \multicolumn{1}{p{3cm}}{sc = 2xx, ct = (audio OR video)} & \multicolumn{1}{p{3cm}}{sc = 4xx OR 5xx ct $\ne$ (audio OR video)} & \multicolumn{1}{p{3cm}}{\texttt{video/audio src=URL}} & \multicolumn{1}{l}{error.message} & \Y & \N & \N\\

		OP-XfoObject & \multicolumn{1}{p{3cm}}{sc =  200, xfo is disabled, ct = text/*} & \multicolumn{1}{p{3cm}}{sc =  200, xfo is enabled} & \multicolumn{1}{p{3cm}}{\texttt{object data=URL}} & \multicolumn{1}{l}{contentDocument} & \Y & \N & \N \\
		
		OP-XfoIFrame & \multicolumn{1}{p{3cm}}{xfo is disabled} & \multicolumn{1}{p{3cm}}{sc = (2xx OR 3xx OR 4xx OR 5xx), xfo is enabled} & \multicolumn{1}{p{3cm}}{\texttt{iframe src=URL}} & \multicolumn{1}{l}{contentDocument} & \Y & \N & \N\\
		
		OP-WindowProperties & \multicolumn{1}{p{3cm}}{sc = 200, ct = text/html, bdy = HTML with window property \textit{A}} & \multicolumn{1}{p{3cm}}{sc = 200, ct = text/html, bdy = HTML with window property \textit{B}}& \multicolumn{1}{p{3cm}}{\texttt{window.open(), (\texttt{form}, \texttt{iframe})}} & \multicolumn{1}{l}{frames.length} & \Y & \Y & \Y\\
		
		postMessage & \multicolumn{1}{p{3cm}}{bdy = postmsg \textit{A} broadcast} & \multicolumn{1}{p{3cm}}{bdy = (postmsg \textit{B} broadcast OR no postmsgs broadcast)} & \multicolumn{1}{l}{\texttt{iframe}, \texttt{window.open()}} & receiveMessage() & \Y & \Y & \Y \\
		
		CSSPropRead & \multicolumn{1}{p{3cm}}{sc = 200, ct = text/css, bdy = CSS with rule A}& \multicolumn{1}{p{3cm}}{sc = 200, ct = text/css, bdy = CSS with rule B}& \texttt{link rel=stylesheet href=URL} & \multicolumn{1}{l}{window.getComputedStyle()} & \Y & \Y & \Y\\
		
		JSError &\multicolumn{1}{p{3cm}}{sc = 200, ct = text/javascript, bdy = JS with \textit{A} no. of errors} & \multicolumn{1}{p{3cm}}{sc = 200, ct = text/javascript, bdy = JS with \textit{B} no. of errors} & \texttt{script src=URL} & \multicolumn{1}{l}{window.onerror()} & \Y & \Y & \Y\\
		
		JSObjectRead & \multicolumn{1}{p{3cm}}{sc = 200, ct = text/javascript, bdy = JS with readable object \textit{A}} & \multicolumn{1}{p{3cm}}{sc = 200, ct = text/javascript, bdy = JS with readable object \textit{B}} & \texttt{script src=URL} & \multicolumn{1}{p{1.5cm}}{window.hasOwnProperty(), prototype tampering, global API redefinition} & \Y & \Y & \Y\\
		
		CSPViolation & \multicolumn{1}{p{3cm}}{sc = 3xx, Location = same origin}& \multicolumn{1}{p{3cm}}{sc = 3xx, Location = different origin} & \multicolumn{1}{p{3cm}}{\texttt{iframe}, \texttt{frame}, \texttt{embed}, \texttt{applet}, \texttt{video}, \texttt{audio}, \texttt{object}, \texttt{link}, \texttt{script}} & \multicolumn{1}{l}{\{``csp- report'':\}} & \Y & \Y & \Y\\
		
		AppCacheError & \multicolumn{1}{p{3cm}}{sc = 200} & \multicolumn{1}{p{3cm}}{sc = (3xx OR 4xx OR 5xx)} & \multicolumn{1}{p{1.8cm}}{\texttt{html manifest=MANIFEST.appcache}} & \multicolumn{1}{l}{AppCache error} & \N & \Y & \N\\
		
		Timing & \multicolumn{1}{p{3cm}}{Load/Resp./Parse time \textit{A}} & \multicolumn{1}{p{3cm}}{Load/Resp./Parse time \textit{B}} & \multicolumn{1}{p{1.8cm}}{\texttt{script}, \texttt{video}, \texttt{img}, \texttt{XmlHttpRequest}...} & \multicolumn{1}{l}{timing side-channel} & \Y & \Y & \Y\\
		\bottomrule
	\end{tabular}
\end{table*}

\subsection{Attack Classes Description}
\label{sec:classDescription}
Table~\ref{tab:classes} details the \attackClasses 
identified by the above process.
For each attack class, the table shows
the name we assigned to the class; 
a description of the two different responses by a \sURL 
that can be targeted using this attack class; 
the attack page logic with 
the methods that can be used to include the \sURL and 
the \xsleak to distinguish the responses;
and the affected browsers. 
In each response description we abbreviate HTTP fields as follows:
Status Code (sc), 
Content-Type (ct), 
X-Content-Type-Options (xcto), 
Content-Disposition (cd), and 
response body (bdy).

\paragraph{EventsFired.}
The first \numEFAttackClasses attack classes use the events fired in the browser
as \xsleak and hence are denoted by the prefix \textit{EF-}.
The first attack class \textit{EF-StatusErrorScript} 
can target \sURLs that return in one state a success status code 
($sc = 200$) with JavaScript (JS) content ($ct = text/javascript$), 
and return an error ($sc = (4xx\, OR\, 5xx)$) in another state. 
The events fired by both types of responses are different 
(onload in one case, onerror in the other) allowing to distinguish 
the two responses.
This attack class works on all browsers.
Among these \numEFAttackClasses attack classes, 
\numNewEFAttackClasses are new and for the other \numKnownEFAttackClasses attack instances had been previously proposed.
Most of these \numEFAttackClasses involve the type or disposition of 
the content, including content-sniffing (\texttt{X-Content-Type-Options}).
There are also cases related to the \texttt{X-Frame-Options} header. 

\paragraph{Object Properties.} 
The next \numOPAttackClasses attack classes leverage as \xsleak 
the readable properties of the included resource. 
Out of these \numOPAttackClasses, \numNewOPAttackClasses are new variations.
For instance, in \textit{OP-ImgDimension}, if a \sURL returns 
images with different dimensions, 
the \textit{height} and \textit{width} properties allow  
to differentiate the responses.
While these two properties were known to leak~\cite{XsleaksCollection2019}, 
our approach uncovers that similar attacks exist using the 
\textit{naturalHeight} and \textit{naturalWidth} properties. 
Interestingly, \textit{OP-ImgCtMismatch} presents a similar attack targeting 
\sURLs that return an image and a non-image, 
which works because for non-image resources 
some browsers return the height and width of a broken image icon, 
triggering a difference in dimensions.
The term (\texttt{form}, \texttt{iframe}) in classes
\textit{OP-FrameCount, OP-WindowProperties} 
captures that it is also possible to include the resource using a 
\texttt{form} tag (using the \texttt{action} attribute) to trigger a 
POST request (specifying \texttt{method} as POST), 
and embedding the response in an iframe
(pointing \texttt{target} attribute to an iframe)~\cite{Evanssearchtiming2009}. 
All other attack classes leverage GET requests.

\paragraph{PostMessage.}
This class uses a novel \xsleak that as far as we know has not been 
previously mentioned. 
It can target \sURLs that return different broadcasted postMessages, 
or a broadcast postMessage and no broadcast. 
It affects all three browsers.
To read the postMessages, the attack page can include the 
\sURL using the \textit{iframe} tag
if the page does not use framing protection, or 
the \textit{window.open} method if framing protection is used.
To identify a difference between responses, it compares 
the number of broadcast messages, 
the message origins, and the message content.
The message content is compared using the 
Jaro string distance~\cite{jaroSimilarity} 
to account for small session-specific or user-specific differences.

\paragraph{CSSPropRead.} 
Another \xsleak leverages \sURLs that return different CSS rules 
for different states. 
To identify the differences, the attack page is designed to contain 
elements affected by the differing rules and to check the inherited style rules.
Some attack instances in this class were previously 
known~\cite{Grossmancoldmany2012,Evanscsslo2008}.
This class complements the \textit{OP-LinkSheet} and 
\textit{OP-LinkSheetStatusError} classes, which can differentiate 
between CSS and non-CSS responses. 

\paragraph{JSError.}
When a \sURL returns different JavaScript files, where one contains a JS error 
and the other does not, this difference can be detected using the 
\textit{window.onerror()} callback function.
The original attack instance used \textit{window.onerror()} to read the 
line number and the type of JS error triggered~\cite{Grossmancosijserror2006}. 
But, since Cross-Site Script Inclusion (XSSI) 
attacks~\cite{Grossmanxssi2006,Takeshixssi2015} 
abused the verbosity of \textit{window.onerror()}, 
popular browsers no longer return the error line. 
However, we find the attack still works by comparing the number of 
errors triggered.
This class complements \textit{EF-StatusErrorIFrame}, 
which allows differentiating JS and non-JS responses. 

\paragraph{JSObjectRead.}
Another \xsleak for differentiating responses that contain JS files
checks the presence or absence of certain readable objects in the included JS.
The original attack instance checked for 
global variables~\cite{Grossmanxssi2006}, but later attacks also leveraged 
techniques such as prototype tampering and global 
API redefinition~\cite{LekiesXSSI2015SEC}.

\paragraph{CSPViolation.}
When a \sURL redirects visitors to the same origin 
in a state and to a different origin in another state, 
this difference can be detected using a Content Security Policy (CSP). 
The attacker configures its attack site with a CSP policy for the attack page
that states that any attempt to load a resource from an origin 
different than the attack site should send a violation report 
back to the attack site.
This method was originally proposed for leaking sensitive information 
in the CSP report (e.g., in the path and subdomain)~\cite{HomakovCSPbug2013}. 
Browsers then removed the path information from CSP reports, but the 
attack still works by focusing on whether the CSP violation report is received
(redirection to different origin) or not (redirection to same origin).

\paragraph{AppCacheError.}
When a \sURL returns a success status code (2xx) in one state and a 
redirection (3xx) or error (4xx, 5xx) in another, 
this difference can be detected through the browser's 
AppCache~\cite{AppCache2014}. 
The attack page uses the {\tt manifest} attribute of the {\tt html} tag
to refer to an AppCache manifest file, 
which includes the \sURL in the list of URLs that should be cached.
This forces the browser to request the \sURL. 
If the \sURL returns a success status code, 
an AppCache {\it cached} event is triggered. 
If the \sURL returns a redirection or error, 
an AppCache {\it error} event is triggered instead. 
Lee et al.~\cite{LeeAppCacheErr2015NDSS} first presented this attack 
showing that it affected five browsers. 
However, this \xsleak currently only works in Chromium-based browsers because  
Firefox and Edge no longer allow cross-origin URLs to be cached using AppCache. 

\paragraph{Timing.} 
Multiple works have shown that timing differences when a resource is 
requested from different states can be used to distinguish 
those states~\cite{BortzTiming2007WWW,Evanssearchtiming2009,GelernterCSSA2015CCS,VanGoethemClock2015CCS,SanchezBakingTimer2019ACSAC}.
Those works focus on acquiring accurate timing 
information resistant to changes in network conditions. 
We have incorporated into \tool the ability to gather accurate 
timing information using the video parsing 
leak in~\cite{VanGoethemClock2015CCS}.

\section{\Tool}
\label{sec:tool}

We have designed and implemented \tool, 
a tool for assisting a security analyst in 
identifying, and generating evidence of, COSI attacks in a target site.
\Tool focuses on the COSI attack preparation phase.
It takes as input a target site, 
a set of state scripts defining states in the target site, and 
the attack classes identified in Section~\ref{sec:classes}.
It outputs attack pages,
which can be used by a security analyst 
for demonstrating the existence of complex COSI attacks,
involving more than two states and supporting multiple browsers.

\paragraph{Setup.}
\Tool needs network access to the target site, 
which may be a local installation of an open-source web application 
(e.g., GitLab, HotCRP) or a remote web site (e.g., \url{linkedin.com}, \url{facebook.com}). 
The analyst needs to be able to create user accounts in the target site.
Those accounts should cover different account types 
and should be populated with content, 
e.g., filling the user profile, creating a blog, adding blog entries.
For example, to test the open source HotCRP conference management system,
the analyst prepares a local installation by creating a test conference and 
five user accounts: 
administrator,
two authors, and 
two reviewers. 
Then, he submits a paper using each of the author accounts. 
Finally, 
it assigns the paper submitted by the first author to the first reviewer and 
the paper submitted by the second author to the second reviewer.

Once the target site is configured, the analyst creates state scripts 
that can be executed to automatically load a specific state at a web browser,  
i.e., to log into the tested web application using one of the created accounts
or to log out of an account. 
\Tool currently supports state scripts written using the 
Python Selenium WebDriver~\cite{SeleniumWebDriverPython}.
The web browser to be used is an argument to the state script. 
In our HotCRP example, the analyst creates six state scripts.
The first five scripts open a web browser, visit the login page, and
authenticate using one of the created accounts. 
The last script logs in and then logs out to capture the logged out state.

\begin{figure}[t]
	\centering
	\caption{\tool architecture. 
  }
	\includegraphics[width=\columnwidth]{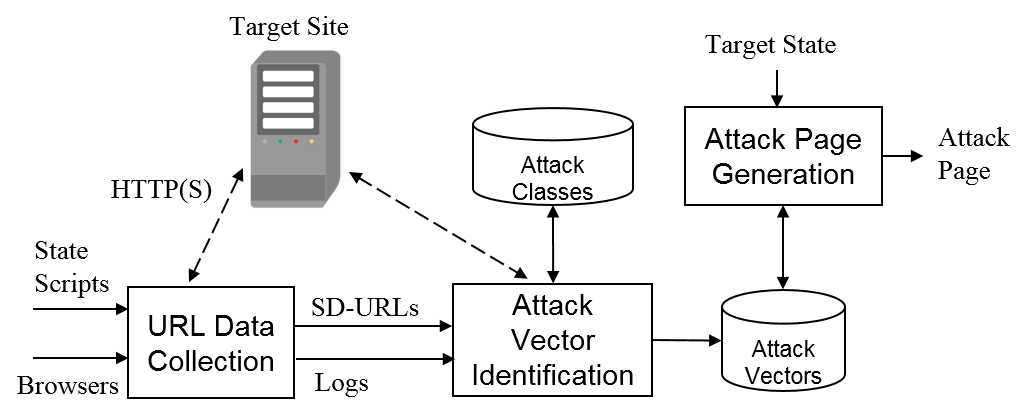}
	\label{fig:arch}
\end{figure}

\paragraph{Architecture.}
The architecture of \tool is shown in Figure~\ref{fig:arch}.
It takes as input the state scripts, a set of browsers, 
the configured target site, and a target state. 
It outputs an attack page that leaks if a victim is in 
the target state at the target site.
\Tool comprises of three modules: 
{\it URL data collection}, 
{\it attack vector identification}, and 
{\it attack page generation}. 

The URL data collection module crawls the target site to discover URLs.
It visits each discovered URL 
to collect its response when visited 
from a specific state with a specific browser.
And, it compares the responses to the same URL obtained from 
different states to identify \sURLs that may be candidates to be used 
in attack pages. 
 
Next, the attack vector identification checks if any of the \sURLs can be 
attacked using the known COSI attack classes. 
When needed, it visits each \sURL using a set of inclusion 
vectors to collect browser events 
that can only be obtained with a specific inclusion method 
(e.g., postMessages), or 
that cannot be easily obtained statically from the HTTP(S) responses
(e.g., JS errors, readable JS objects). 
For each \sURL that matches an attack class, it outputs an attack vector. 

Finally, the attack page generation module 
builds an attack page that enables identifying if the victim is in the target 
state at the target site. 
The generated attack page may combine multiple attack vectors to 
uniquely distinguish the target state and to support multiple browsers. 
Attack pages for different target states can be created by re-running the 
attack page generation module, without re-running the previous modules.

\begin{table*}[t]
	\caption{Examples of URLs collected from HotCRP from three states. 
		For simplicity, the response is represented with only a 
		subset of 4 field values:
		Status Code (sc), Content-Type (ct), X-Frame-Options (xfo), and
		X-Content-Type-Options (xcto).}
	\label{table:HotCRPurls}
	\centering
	\scriptsize	
	\begin{tabular}{llllll}\toprule
		\textbf{URL} & \multicolumn{5}{p{10cm}}{\centering \textbf{Response Received at Different States}} \\
		& \multicolumn{1}{p{3cm}}{\textit{Reviewer1 (R1)}} & 
		\multicolumn{1}{p{3cm}} {\textit{Reviewer2 (R2)}} & 
		\multicolumn{1}{p{3cm}} {\textit{Logged Out (LO)}}
		\\ \hline
		\multicolumn{1}{l}{\url{/testconf/images/pdffx.png}}  & 
		\multicolumn{1}{p{3cm}}{sc = 200, ct = image/png, no xfo, no xcto} & 
		\multicolumn{1}{p{3cm}}{sc = 200, ct = image/png, no xfo, no xcto} &
		\multicolumn{1}{p{3cm}}{ sc = 200, ct = image/png, no xfo, no xcto} \\
		\multicolumn{1}{l}{\url{/testconf/api.php/review?p=1}}  & 
		\multicolumn{1}{p{3cm}}{sc = 200, ct = text/html, no xfo, xcto = nosniff} & 
		\multicolumn{1}{p{3cm}}{sc = 403, ct = text/html, no xfo, no xcto} &
		\multicolumn{1}{p{3cm}}{ sc = 200, ct = text/html, no xfo, no xcto} \\
		\multicolumn{1}{l}{\url{/testconf/offline.php?downloadForm=1}}  & 
		\multicolumn{1}{p{3cm}}{sc = 200, ct = text/html, no xfo, xcto = nosniff} & 
		\multicolumn{1}{p{3cm}}{sc = 200, ct = text/html, no xfo, xcto = nosniff} &
		\multicolumn{1}{p{3cm}}{sc = 200, ct = text/html, no xfo, no xcto} \\
		
		\hline
	\end{tabular}
\end{table*}

\subsection{URL Data Collection}
\label{sec:crawler}

The URL data collection module performs three main tasks: 
crawling to discover URLs, 
collecting the responses for each URL 
when visited from a specific state with a specific browser, and 
identifying \sURLs.
The module is built on top of the Spider crawler 
for OWASP ZAP~\cite{zap}. 
The crawling considers a URL to be part of the target site 
if it satisfies at least one of three constraints:
it is hosted at the target site domain,  
it redirects to a URL hosted at the target site domain, or 
it is part of a redirection chain involving a URL satisfying any of the above 
two criterion.

Each discovered URL is visited from each input state and using each 
input browser.
Before visiting a URL, a state script is executed to load 
the corresponding state in the browser.
The state scripts also allow collecting URLs only accessible 
from authenticated states.  
Currently, \tool supports the three most popular browsers: 
Chrome, Firefox, and Edge.
For each browser, it supports the latest version at the time 
we started the implementation: 
Google Chrome 71.0.3578.98, 
Mozilla Firefox 65.0.1, and 
Microsoft Edge 42.17134.1.0. 
The module has a flexible design that allows adding support for other browsers 
and browser versions.
For each triplet (URL, browser, state), 
it stores the full response (headers and body) received from the server. 
URLs that return the same response in each state are not state-dependent 
and thus cannot be used in a COSI attack. 
To identify if a URL is state-dependent, 
a similarity function is used that compares responses 
ignoring non-deterministic fields such as the {\tt Date} header or CSRF 
tokens that may differ in each response.
URLs that return the same response (minus non-deterministic fields) 
in every state are not state-dependent, and can be discarded.

To illustrate the tool we use our HotCRP running example 
with only three state scripts: 
Reviewer1 (R1), Reviewer2 (R2), and LoggedOut (LO).
The goal of the analyst is to find a COSI attack that  
reveals the reviewer of a specific paper. 
In this scenario, the tester can ignore the administrator and author accounts 
since an attacker (typically an author) would only send emails with the 
attack page URL to the (non-chair) PC members.
The three identified URLs in our running example are shown in 
Table \ref{table:HotCRPurls}.
Each table entry shows the response for the URL when visited from a 
specific state. 
For simplicity, each response is summarized as a tuple of 4 field values:
Status Code (sc), Content-Type (ct), X-Frame-Options (xfo), and 
X-Content-Type-Options (xcto). 
The URL \url{/images/pdffx.png} is not a \sURL since it returns 
the same response in all states. 
Thus, it will be removed at this step. 
The other two URLs are state-dependent since for each of them there
exists at least one pair of states whose responses are different. 

\subsection{Attack Vector Identification}
\label{sec:sdurls}

The goal of the attack vector identification module is to find, 
among all the \sURLs discovered, the ones for which a matching 
attack class is known, 
and thus can be used to generate attack vectors.
\Tool supports all attack classes in Table~\ref{tab:attacks}. 
Those attack classes can be split into two groups. 
The first (static) group are attack classes for which it can be determined,
using solely the collected logs of HTTP(S) responses, 
if a \sURL matches the class. 
This group includes all classes that capture differences in HTTP headers 
such as Status Code, Content-Type, or X-Frame-Options.
The second (dynamic) group are attack classes for which matching a \sURL 
requires data difficult to obtain from the responses 
such as JS errors, postMessages, and audio/video properties 
(e.g., width, height, duration).
For this group, it is needed to visit the \sURL with 
different inclusion methods to collect the missing data.

For each \sURL and pair of states that return different responses for that 
\sURL, the module first checks if there exist any 
matching static attack classes. 
For efficiency,
if two different state pairs produce the same responses, there is no 
need to query the attack classes for the second pair.
We illustrate this process using the \sURLs in Table \ref{table:HotCRPurls}.
For \texttt{api.php}, the responses from (R1, R2) match two static attack classes: 
\textit{EF-StatusErrorObject} (for Firefox and Edge), 
\textit{EF-StatusErrorLink} (for Chrome).
Similarly, the responses from (R2, LO) match the same two static attack classes as (R1, R2).
Finally, the states (R1, LO) 
match the static attack classes \textit{EF-XctoObject} and \textit{EF-XctoScript}. 
The process repeats with the other \sURL (\texttt{offline.php}).
Since states R1 and R2 return the same response, 
(R1, R2) can be ignored. 
For states (R1, LO), the attack classes \textit{EF-XctoObject} and \textit{EF-XctoScript} match.
Finally, for states (R2, LO) the responses are the same as for (R1, LO) and 
there is no need to check them again.

In our example, all state pairs can be distinguished using a static attack class. 
If that was not the case, the module would collect additional information to 
check the dynamic attack classes.
For this, the \sURL is included in a set of data collection pages 
hosted at a test web server.
Each page uses an inclusion method from one of the dynamic classes and 
collects the required dynamic data for the class
(e.g., use \textit{script} to collect JS errors and JS readable objects). 
Each data collection page is visited with each browser and from every state 
that returns a unique response.

The attack vector identification module outputs, 
for each pair of states, a list of pairs (\sURL, AttackClass) 
specifying that an attack vector that uses the \sURL and the attack class 
can distinguish those two states for the browsers defined by the attack class.

\subsection{Attack Page Generation}
\label{sec:attackpage}

Given a target state $s_t$ and a set of target browsers $B$, 
the goal of the attack page generation is to produce an attack page 
that combines attack vectors to uniquely distinguish $s_t$ 
from the other states, when visited by a browser in $B$. 
The set of target browsers should be equal to or a subset of the 
set of browsers input to \tool. 
This process comprises of two steps: 
attack vector selection and attack page construction.

\normalem
\IncMargin{1em}
\begin{algorithm}[t]
	\SetAlgoLined
	\scriptsize
	\SetKwInOut{Input}{inputs}
	\SetKwInOut{Output}{outputs}
	\SetKwFunction{filter}{filter}
	\SetKwFunction{merge}{mergeStates}
	\SetKwFunction{score}{score}
	\SetKwFunction{getSupportedSBPermuations}{getCoveredPairs}
	
	\Input{Target state $s_t$, target browsers $B$, states $S$, attack vectors $A$ }
	\Output{The list of selected attack vectors}
	\BlankLine
	outVectors $\leftarrow$ [ ]\;
	$S_r \leftarrow S - s_t$\;
	$A_r$ $\leftarrow$ \filter{$A$, $s_t$}\;
	$A_r$  $\leftarrow$ \merge{$A_r$}\;
  $P \leftarrow$ (s\textsubscript{i} $\in S_r$,  b\textsubscript{j} $\in$ $B$)\;
  \While {$P \ne \emptyset, A_r \ne \emptyset, s > 0$}{
    V = \score{$A_r,P$}\;
    (s,a) $\leftarrow$ (max(V),argmax(V))\;
		\If{$s > 0$}{
      outVectors.append(a)\;
      $P \leftarrow$ $P$ - \getSupportedSBPermuations{a}\;
      $A_r \leftarrow A_r - a$\;
    }
  }
	\textnormal{\textbf{return}} outVectors, $P$ \;
	\caption{Attack vector selection}
	\label{alg:avs}
\end{algorithm}\DecMargin{1em}

\ULforem

Algorithm~\ref{alg:avs} details the attack vector selection.
It selects, among all attack vectors, 
the ones needed to distinguish the target state when visited by a 
target browser. 
The algorithm first removes all attack vectors that do not include the 
target state since they do not enable distinguishing $s_t$ (Line 3).
In our HotCRP example, the target state is R1 and 
all attack vectors for state pair (R2, LO) are removed.
Then, it merges the states of all remaining attack vectors with 
the same \sURL and attack class into a single attack vector 
that distinguishes $S_t$ from $n \ge 2$ other states. 
In our example, the attack vectors do not merge further.
Next, it initializes a set $P$ with all pairs of states and browsers to 
be distinguished (Line 5).
The algorithm goes into a loop that at each iteration it identifies the 
attack vector that covers most remaining pairs in $P$ (Lines 6-14). 
The loop iterates until all pairs have been covered, 
no attack vectors remain, or the remaining attack vectors do not allow 
distinguishing the remaining pairs.
To select an attack vector, a \texttt{score} function is used that 
assigns higher scores to attack vectors that cover more pairs in $P$, 
penalizing attack classes that may interfere with other vectors (Line 7). 
For example, an \textit{EventsFired} attack vector using the \texttt{script}
tag may trigger CSP violation reports that interfere with a CSP policy 
for \textit{CSPViolation} that targets script resources.
If the score is zero, the loop breaks as the remaining attack vectors 
do not allow distinguishing the remaining pairs. 
Otherwise, the selected attack vector is appended to the output (Line 10), 
the newly covered pairs are removed from $P$ (Line 11), and 
the attack vector is removed from the available list (Line 12).

In our example, the first loop iteration selects the attack vector 
(\{LO\}, \texttt{offline.php}, \textit{EF-XctoObject})
as it covers three pairs, differentiating the logout state for Chrome, Firefox and Edge.
The next loop iteration selects the attack vector 
(\{R2\}, \texttt{api.php}, \textit{EF-StatusErrorObject}) as it covers two other pairs,
differentiating all remaining states for Firefox and Edge.
Finally, the last iteration chooses (\{R2\}, \texttt{api.php}, \textit{EF-StatusErrorLink})
which covers the remaining state for Chrome. 
At that point, no more pairs remain to be covered, and the algorithm 
outputs the selected attack vectors.
The algorithm also outputs the pair set $P$. 
If empty, the attack page distinguishes the target state from 
all other states for all target browsers. 
Otherwise, some states may not be distinguishable for some target browsers.

For each attack class, the attack page generation module has a template 
to implement the attack. 
For each selected attack vector, it chooses one inclusion method in the 
attack class, and applies the corresponding template with the \sURL.
All instantiated templates are integrated into the output attack page.
\section{Ethics}
\label{sec:ethics}

Our experiments do not target any real user of the live sites.
All testing on live sites is restricted to user accounts that we
created on those sites exclusively for this purpose.
The process of validating that the attacks found on open-source
web applications work on live installations of those applications
is similarly restricted to accounts owned by the authors.
The impact on live sites is limited to receiving a few thousand 
requests for valid resources in the site.
We take two actions to limit the load on live sites from our testing. 
First, we spread the requests over time to avoid spike loads.
Second, we disable the timing \xsleak in our experiments, 
which requires sending hundreds, or even thousands, of requests per \sURL,
generating the highest load. 

We have disclosed our attacks to the four web applications,
receiving confirmation of the issues from HotCRP, GitLab, and GitHub,
while OpenCart has not replied.
The disclosure process for the web sites is ongoing.
All reported attacks have been confirmed and
some attacks have already been patched (e.g., HotCRP, \url{linkedin.com}).
We avoid providing \sURLs for attacks not yet patched.
We have also reported our results to the three browser vendors, 
as well as the Tor project.
We incorporate their feedback into our defenses
discussion in Section~\ref{sec:defenses}.

We acknowledge that publicly releasing \tool makes it possible for
attackers to misuse it to find COSI attacks.
However, we argue that this applies to any penetration testing and
vulnerability discovery tool (open source or commercial).
Other distribution models such as Software-as-a-Service could potentially
mitigate this risk, but would also limit the usefulness for the
research community.
We believe determined attackers will still find a way to attack sites
even without \tool.
Thus, we favor the benefit for defenders and the research community.
\section{Experiments}
\label{sec:experiments}

This section presents the evaluation of \tool 
on four open source web applications 
(HotCRP, GitLab, GitHub Enterprise, OpenCart) and 
the 58 web sites in the Alexa Top 150~\cite{AlexaTop500}
where we could create user accounts.
These targets are popular, 
allow us to test on white-box (open source) and black-box (deployed) scenarios, 
and cover services with multiple user states. 
Section~\ref{sec:evalApps} describes the results on Web applications, 
Section~\ref{sec:evalAlexa} on Alexa web sites, 
and Section~\ref{sec:evalCases} details some attacks found. 

\begin{table*}
  \caption{\Tool evaluation results.
  For every target application and site, 
  it shows the data for each tool module, 
  as well as the type and browsers affected for the 
  attacks found. 
  Browsers are abbreviated as Chrome (C), Firefox (F), and Edge (E).}
  \label{tab:results-overview}
  \centering
  \scriptsize
  \renewcommand{\arraystretch}{1.2}
  \begin{tabular}{l|rrr|rrr|rrrrr| llll}\toprule
  	\multicolumn{1}{l|}{} & 
    \multicolumn{3}{c|}{\textbf{Data Collection}} & 
    \multicolumn{3}{c|}{\textbf{Attack Vector Identification}} &
    \multicolumn{5}{c|}{\textbf{Attack Page Generation}} &
    \multicolumn{4}{c}{\textbf{Attacks Found}} \\
    \cline{2-16}
    \multicolumn{1}{l|}{\textbf{Target}} & 
    \multicolumn{1}{c}{\textbf{}} & 
    \multicolumn{1}{c}{\textbf{}} & 
    \multicolumn{1}{c|}{\textbf{SD}} & 
    \multicolumn{1}{c}{\textbf{}} & 
    \multicolumn{1}{c}{\textbf{State}} & 
    \multicolumn{1}{c|}{\textbf{}} &
    \multicolumn{1}{c}{\textbf{UD}} &
    \multicolumn{1}{c}{\textbf{PD}} &
    \multicolumn{3}{c|}{\textbf{Vectors}} &
    \multicolumn{1}{l}{\textbf{Login}} & 
    \multicolumn{1}{l}{\textbf{Account}} & 
    \multicolumn{1}{c}{\textbf{}} & 
    \multicolumn{1}{l}{\textbf{Access}} \\ 
    & \textbf{States}
    & \textbf{URLs}
    & \textbf{URLs}
    & \textbf{Vectors}
    & \textbf{Pairs} 
    & \textbf{\xsleaks}
    & \textbf{States} 
    & \textbf{States} 
    & \textbf{Min} 
    & \textbf{Avg} 
    & \textbf{Max} 
    & \textbf{Detection} 
    & \textbf{Type} 
    & \textbf{Deanon.} 
    & \textbf{Detection} \\
    \hline
    HotCRP&5&68&65&116&7&3&1&4&1&1.6&3&\multicolumn{1}{l}{C,E,F}&\multicolumn{1}{l}{-}&\multicolumn{1}{l}{C,E,F}& \multicolumn{1}{l}{-}\\ 
    
    GitLab&6&52&19&236&14&1&2&4&1&1.9&2&\multicolumn{1}{l}{C,E,F}&\multicolumn{1}{l}{C,E,F}&\multicolumn{1}{l}{C,E,F}&\multicolumn{1}{l}{-}\\
    
    GitHub&4&91&90&992&6&1&4&0&1&1.8&2&\multicolumn{1}{l}{C,E,F}&\multicolumn{1}{l}{C,E,F}&\multicolumn{1}{l}{C,E,F}&\multicolumn{1}{l}{-}\\ 
    
    OpenCart&5&51&32&72&7&1&2&3&1&1.1&2&\multicolumn{1}{l}{C,E,F}&\multicolumn{1}{l}{-}&\multicolumn{1}{l}{-}&\multicolumn{1}{l}{-}\\ 
      
    \midrule 
  \end{tabular}
\end{table*}

\subsection{Evaluation on Web Applications}
\label{sec:evalApps}

Table~\ref{tab:results-overview} summarizes the results of applying 
\tool on the four web applications we installed locally.
It details the results for each tool module,
as well as the COSI attacks found.
The data collection part shows 
the number of input state scripts provided to \tool, 
the number of URLs crawled, and 
the number of \sURLs identified.
The attack vector identification part shows 
the total number of attack vectors identified, 
the number of state pairs they cover, and 
the number of \xsleaks they use.
The attack page generation part shows 
the number of states uniquely distinguished (UD) from other states, 
the number of states partially distinguished (PD) 
excluding UD states,
and the minimum/average/maximum attack vectors in the attack pages. 
Finally, the attacks found part shows the type and browsers affected for the 
identified attacks.

Depending on the target, we created 3--6 state scripts to use \tool. 
One script always corresponds to the logged out (LO) state and 
the others are target-specific.
For example, for GitLab the other 5 states are for 
maintainer, developer, reporter, guest (read-only access), and
a user with no read access to the repository. 
Like a fuzzing tool, \tool will try to find attacks until 
the allocated time budget runs out. 
We let \tool run for a maximum of 24 hours on each target, 
although after a few hours the crawling typically does not find any new URLs. 
The data collection results show that \sURLs are very common, 
on average 
68\%
of the discovered URLs are \sURLs
(and up to 99\% in GitHub). 

\Tool finds between 58 and 992 attack vectors in each target 
using up to 3 \xsleaks. 
The results show that on average the generated attack pages use 
more than one attack vector. 
Account type and deanonymization attacks always require multiple vectors, 
while login detection is oftentimes possible with a single vector.
This highlights the importance of our approach to combine attack vectors 
in order to handle more than two states and multiple browsers.
Some states can be uniquely identified, 
i.e., distinguished from any other state, and 
the rest can be partially distinguished. 
We found no state that could not be distinguished at all.
It is important to note that partially distinguishable states can also
be used in attacks. 
For example, not being able to differentiate the administrator 
from a normal user does not matter if the administrator is not targeted 
by the attack, i.e., not sent the attack page URL.
Overall, \tool finds attacks on all four applications: 
login detection attacks on all four, 
deanonymization attacks on three, and account type identification on two.

\begin{table}
	\caption{Web sites vulnerable to each attack type}
	\label{table:overall-vuln-dst}
	\centering
	\scriptsize
	\renewcommand{\arraystretch}{1.2}
	\begin{tabular}{rrr}\toprule
		\multicolumn{1}{l}{\textbf{Attack Type}} & 
		\multicolumn{1}{l}{\textbf{Tested}} &
		\multicolumn{1}{l}{\textbf{Vulnerable}} \\ \midrule
		\multicolumn{1}{l}{Login Detection} & 58 & 58 \\
		\multicolumn{1}{l}{Deanonymization} & 58 & \nUserIdentAtks \\
		\multicolumn{1}{l}{SSO Status} & 12 & 12\\
		\multicolumn{1}{l}{Access Detection} & 11& 5\\
		\multicolumn{1}{l}{Account Type} & 3 & 3\\
		\hline
		\multicolumn{1}{l}{\textbf{Total Sites}} & 58 & 58\\
	\end{tabular}
\end{table}

\begin{table}
  \vspace{-6pt}
	\caption{Attack vectors found per \xsleak and browser.}
	\label{table:overall-atk_type-dist}
	\centering
	\scriptsize
	\renewcommand{\arraystretch}{1.2}
	\begin{tabular}{llrrrrrrrrr}\toprule
		\multicolumn{1}{l}{\textbf{Attack}} & 
		\multicolumn{1}{l}{\textbf{Br}} &
		\multicolumn{1}{c}{\textbf{EF}} & 
		\multicolumn{1}{c}{\textbf{OP}} & 
		\multicolumn{1}{c}{\textbf{PM}} & 
		\multicolumn{1}{c}{\textbf{CSS}} & 
		\multicolumn{1}{c}{\textbf{JSE}} &
		\multicolumn{1}{c}{\textbf{JOR}} &
		\multicolumn{1}{c}{\textbf{CSP}} &
		\multicolumn{1}{c}{\textbf{ACE}} 
		\\ \hline

		\multirow{3}{.7cm}{Login Detect.}&C&2457&2532&9&0&2&0&885&63\\
		&F&1587&1511&9&0&2&0&424&0\\
		&E&676&1286&9&0&2&0&434&0\\ \midrule

		\multirow{3}{.7cm}{Account Type}&C&175&82&0&0&0&0&126&3\\
		&F&173&85&0&0&0&0&2&0\\
		&E&39&36&0&0&0&0&12&0\\ \midrule

		\multirow{3}{.7cm}{Deanon.}&C&644&546&2&0&0&0&31&17\\
		&F&447&420&2&0&0&0&79&0\\
		&E&201&288&2&0&0&0&81&0\\ \midrule
		
		\multirow{3}{.7cm}{Access Detect.}&C&98&12&0&0&0&72&0&0\\
		&F&1&10&0&0&0&0&0&0\\
		&E&3&10&0&0&0&0&0&0\\ \midrule
		
		\multirow{3}{.7cm}{SSO Status}&C&0&0&0&0&0&0&12&0\\
		&F&0&0&0&0&0&0&12&0\\
		&E&0&0&0&0&0&0&12&0\\ 

		\midrule
	\end{tabular}
	\newline (\textbf{Legend:} EF=EventFire; OP=ObjectProperties; PM=PostMessage; CSS= CSSPropRead; JSE=JSError; JOR=JSObjectRead; CSP=CSPViolation; ACE=AppCacheError)
\end{table}
\vspace{-12pt}

\subsection{Evaluation on Web Sites}
\label{sec:evalAlexa}

We test sites from the Alexa Top 150 that are not duplicates 
(e.g., \url{amazon.com} vs. \url{amazon.de}) and where we could create free accounts. 
This excludes sites without user accounts, 
that required a phone number in a specific area, 
or that demanded credit card information.
This leaves us with \nSites sites, 
of which only 12 support SSO, and only 3 have multiple types of free accounts 
(excluding the administrator account that we cannot obviously create).
For access detection, we focus on privacy sensitive sites, 
more specifically adult sites, on the Alexa Top 150, 
regardless if they have user accounts. 

Table~\ref{table:overall-vuln-dst} summarizes the number of tested and 
vulnerable sites for each attack type. 
For login detection, SSO status, and account type 
identification, \tool discovers \xsleaks against all tested sites.
In addition, it finds deanonymization attacks 
in 57\% of the sites and access detection attacks in 45\%.
The results show that login detection attacks 
are easiest to find, but that by combining multiple attack vectors it is 
possible to find more powerful attacks targeting more than two states in 
72\% of the sites.
Regarding false positives, we rarely observed them in two situations. 
One was due to \tool waiting 6 seconds to collect events and some 
pages being slower to load. 
The other one was when \tool sent too many queries and a site 
started replying with CAPTCHAs. 
We expect that increasing the timeout and distributing the queries over 
multiple IPs would eliminate those false positives.
We do not evaluate false negatives, as we lack ground truth of the 
COSI attacks present in the targets.
However, we acknowledge that, like any testing tool, 
false negatives are possible, 
e.g., \tool can only find COSI attacks that are instances of the 
\attackClasses it supports.

The support in \tool for multiple \xsleaks and multiple browsers 
allows to compare the prevalence of the \xsleaks, 
as well as the attack surface of the browsers, 
on the same set of \sURLs, 
i.e., independently of the crawler's coverage.
Table~\ref{table:overall-atk_type-dist} details the distribution of 
attack vectors per \xsleak for each attack type and browser pair. 
\xsleak prevalence widely varies. 
Most attack vectors use \textit{EventsFired}, \textit{Object Properties}, and 
\textit{CSPViolation} \xsleaks. 
Our novel \textit{postMessage} \xsleak ranks sixth out of eight \xsleaks, 
producing attack vectors on 11 different sites including 
\url{blogger.com}, \url{ebay.com}, \url{reddit.com}, and \url{youtube.com}.
The least prevalent \xsleak is \textit{CSSPropRead} for which 
\tool does not find any attack vector,
showing that \sURLs on CSS content that leak user state are not common.
The comparison also shows that Chrome has a larger attack surface, 
ranking first in number of attack vectors in all eight \xsleaks.

\subsection{Example Attacks}
\label{sec:evalCases}

This section details some of the attacks \tool found that 
involve more than two states.
All attacks work on the three tested browsers, 
unless specifically noted.

\paragraph{HotCRP.}
\tool found an attack for determining whether the victim is 
a reviewer of a specific paper, which we have used as running example. 
The attack page (Listing~\ref{lst:attackpage}) uses three attack vectors, 
one for login detection on all three browsers, 
and two (one for Chrome and another for Firefox/Edge) 
to identify if the victim submitted a review for the target paper.
To launch the attack, the attacker collects the email addresses of
the program committee members and sends them a spear-phishing email
to convince them to click on the attack page URL.
Since the attack was found on a local HotCRP installation, 
to test it on conferences hosted at \url{hotcrp.com}, 
we had to update the \sURLs with the proper domain and conference name.
We verified the attack and reported it to the HotCRP developer, 
who confirmed the issue and has released a patch~\cite{HotCRPCOSIAck2019}.

\paragraph{GitLab and GitHub.}
Attacks are found in both GitLab and \url{github.com} that allow 
determining if the victim is the owner of a repository 
(or of a snippet). 
Both attacks first use a login detection attack. 
If the victim is logged in, the attack page uses an \textit{EventFire} 
attack class using a \sURL for editing the repository settings (or the snippet) 
to detect if the victim has administrative rights.
For GitHub Enterprise installations, 
another attack allows distinguishing the administrator from other users 
by including the URL for accessing staff tools.

\paragraph{LinkedIn.}
A \textit{CSPViolation} attack allows distinguishing the account type 
(free or premium) using the \sURL \url{https://www.linkedin.com/cap/}.  
This attack has already been fixed following our disclosure.
A second attack allows determining if the victim owns a 
specific LinkedIn profile using the \textit{OP-WindowProperties} attack class. 
The underlying cause of this attack is that the number of frames in a LinkedIn 
profile page is 3 when visited by the owner of the profile, and 4 otherwise.

\paragraph{Blogger.}
Multiple deanonymization attacks are found for 
determining if the victim is the owner of a specific blog. 
The attacker needs to know the \textit{blogID} of the target victim, which can
be found on the HTML source of the target blog.
The attacks combine a \textit{CSPViolation} login detection attack vector
with another deanonymization attack vector from different attack classes 
(e.g., \textit{postMessage},  \textit{EF-CtMismatchScript}). 
This shows how attacks can combine multiple attack vectors 
using different \xsleaks, highlighting the value of our generic approach 
not being specific to any \xsleak. 

 \paragraph{IMDB.}
 A deanonymization attack allows determining if the victim owns a 
 specific IMDB account using a \sURL that contains the user identifier. 
 This attack can determine if the visitor is a specific person from
 the film industry by including the user identifier obtained from the profile 
 for that person.

\paragraph{Amazon.}
\textit{CSPViolation} attacks are found 
that leak if the victim is using the Amazon Kindle Direct Publishing (KDP) 
service, or has accepted the KDP terms and policies.
That information could be used for targeted advertising, 
e.g., to show advertisements of kindle books to the victim. 

\paragraph{Pornhub.}
Attacks are found using the \textit{OP-Window-Properties} and 
\textit{OP-FrameCount} for determining if the victim is the
owner of a specific username, thus enabling deanonymization of the account 
in a closed-world setting. 
The underlying reason for the \textit{OP-FrameCount} attack is 
similar to that of the LinkedIn attack, but mounted on Pornhub's playlist URLs. 

\paragraph{Pinterest.}
A \textit{CSPViolation} attack can be mounted with the 
Facebook SSO initiation URL 
for determining whether the victim authenticated into Pinterest using its 
Facebook account. 
A similar attack was found for Google's SSO.

\paragraph{Imgur.} 
An attack based on \textit{EF-StatusErrorScript} 
can be used to determine if the victim uploaded an image 
(e.g., copyrighted, taken without permission) to this image sharing site. 
The vendor has awarded us a bug bounty for this report~\cite{imgurDiscl2019}.
\section{Defenses Against COSI Attacks}
\label{sec:defenses}

This section discusses existing and upcoming
defenses against COSI attacks.

\paragraph{SameSite cookies.}
\avinash{Post-rebuttal: Mention default same-site protection plan by browsers with porper citation}
COSI attacks leverage the automatic inclusion of 
HTTP cookies~\cite{Barth2011httpcookies}, 
client-side certificates~\cite{Johnsrequestrodeo2006owasp}, and 
HTTP Authentication credentials~\cite{Franks1999httpauth} 
in requests sent by web browsers, 
known as the ambient authority problem in 
browsers~\cite{CzeskisCSRF2013WWW}.
Web sites can use the \texttt{SameSite} attribute in a \texttt{Cookie} header
to prevent the browser from sending that cookie in 
cross-site requests~\cite{Westsamesitecookies2016,JancCOLeaks2018}. 
This defense disables \sURLs whose responses are based on states saved in 
cookies. 
On the other hand, it does not prevent leakage by 
HTTP Authentication credentials and client-side certificates,
it needs to be set for each cookie; 
it may be challenging to deploy in web sites with 
legitimate cross-origin requests~\cite{Sharma2017dropboxsamesite}; and 
its implementation in browsers can have flaws~\cite{Franken3rdPartyCookie2018SEC}.
When we disclosed our results to the browser vendors, 
we were told they plan to address COSI attacks by 
marking all cookies by default as \texttt{SameSite=Lax}, 
unless the site specifically disables them with \texttt{SameSite=None}, 
or makes it stricter with 
\texttt{SameSite=Strict}~\cite{WestSameSiteLaxDefaultIETFDraft2019}.
This change is already planned for 
Chrome~\cite{BlinkSameSiteLaxDefault2019} and 
Firefox~\cite{GeckoSameSiteLaxDefault2019}.
However, this defense will initially ship behind a configuration option 
since it may affect functionality that requires cross-origin requests.

\paragraph{Session-specific URLs.} 
Web sites can use URLs that include a session-specific, non-guessable, token.
The token must be cryptographically bound to the session identifier 
(e.g., the hash of the identifier), 
and the web site must verify this relationship for all HTTP requests. 
Session-specific URLs prevent the attacker from identifying \sURLs 
for the victim's session, avoiding COSI attacks. 
This defense does not depend on browser vendors and can 
be deployed right away.
On the other hand, it can be costly to deploy, 
increases complexity, may impact performance, 
and the web site must ensure that the tokens cannot be leaked or 
brute forced~\cite{CzeskisCSRF2013WWW}.

\paragraph{Cross-Origin-Resource-Policy.}
An emerging HTTP response header that allows web sites to ask browsers to 
disallow cross-origin 
requests to specific resources~\cite{corp}. 
The request is not prevented, rather the browser avoids leakage 
by stripping the response body.
Currently supported by Chrome and Safari.

\paragraph{Fetch metadata.}
An emerging set of HTTP request headers that send additional provenance 
data about a request~\cite{Westfetchmeta2018}, 
e.g., the HTML element triggering a cross-site request.
Currently supported by Chrome.
A web site can use this information to design policies that block 
potentially malicious requests. 
e.g., inclusion of a non-image resource with an \textit{img} tag.

\paragraph{Cross-Origin-Opener-Policy.}
There is ongoing discussion on a new 
HTTP response header to prevent malicious web sites from 
abusing other web sites by opening them in a window~\cite{cowp}. 
This defense could protect against COSI attack classes that 
use the \textit{window.open} inclusion method
(e.g. \textit{OP-Window Properties}, \textit{postMessage}).

\paragraph{Tor Browser.}
\avinash{Post-rebuttal: Provde Tor disclosure details}
The Tor Browser takes preventive measures against 
timing-based COSI attacks~\cite{TorDesignSpec}. 
Additionally, it isolates the browser's state based on the URL in the address bar. 
Therefore, it does not attach cookies and \texttt{Authorization} header values 
to cross-origin HTTP requests generated by inclusions using HTML tags. 
However, the state isolation is not enforced for the \textit{window.open} method, 
so authentication headers are still attached to HTTP requests generated 
using this inclusion method. 
Therefore, Tor Browser users are still vulnerable to \textit{OP-WindowProperties} and
the new postMessage attack class we discovered.

\paragraph{\sURL patching.}
When reporting our attacks, we mentioned SameSite cookies as a good defense
in terms of protection,
since it tackles the root cause of COSI attacks, and cost to deploy. 
However, the developers that already patched our attacks did not take that 
suggestion and instead applied a fix specific to the reported \sURLs. 
For example, the HotCRP developer mentioned that SameSite cookies is 
not available in PHP until PHP 7.3, and instead modified the code to 
always return a 200 HTTP status code with JSON content. 
This fixes our attack, but it will not fix future attacks on other 
status codes and content types. 
In another example, LinkedIn patched our reported user deanonymization 
\textit{OP-FrameCount} attack by making sure that the reported \sURL 
returned the same number of frames for all users.
These examples show that developers currently consider URL-specific fixes a 
quick solution, despite its lack of generality.

\section{Discussion}
\label{sec:discussion}

This section discusses limitations of our approach and 
possible future improvements. 

\paragraph{Preparation overhead.}
To use \tool, the tester first needs to create accounts at the target site 
and provide state scripts that use those accounts. 
Similar overhead is required by other web security testing tools, 
when they need to examine the logged in parts of a web site. 
Furthermore, \tool is designed for web site administrators to test 
their own sites. 
We believe the cost of creating test accounts for your own site is a 
reasonable one-time effort, as these accounts can then be reused 
for other tests. 
In fact, we expect many sites to already have such test accounts 
in place for other types of testing. 

\paragraph{Support for other browsers.}
\tool currently supports the three most popular browsers: 
Chrome, Firefox, and Edge. 
We did not include support for Safari because we run our experiments 
on Windows and Apple stopped releasing Safari for Windows in 2012. 
Adding support for other browsers is a matter of additional 
engineering work. 
Of particular interest would be adding support for mobile platform 
browsers given their popularity and that COSI attacks on those browsers 
have been little explored. 
Support for mobile browsers in \tool could be achieved by integrating a 
mobile testing platform, e.g., Appium~\cite{appium}. 

\paragraph{Support for other crawlers.}
\Tool uses ZAP's Spider module~\cite{Spider} for crawling the target site. 
The coverage of this crawler may be limited on JavaScript-intensive 
web sites. 
It is likely that some \sURLs were not discovered by the crawler 
for this reason, which may have caused COSI attacks to go unnoticed. 
{\Tool}'s modular design should easily allow to integrate other crawlers 
to increase coverage.
Still, despite the potentially limited crawling, 
\tool was able to find COSI attacks in all tested targets.

\paragraph{Dynamic page element detection.}
To identify \sURLs, \tool removes dynamic page elements from HTTP responses.
Our detection of some dynamic page elements, e.g., CSRF tokens, 
is based on heuristics that could introduce errors. 
However, there are a couple of mitigating reasons, 
which may explain why we did not observe such errors in our testing. 
First, even if a URL is wrongly identified as a \sURL, 
\tool may later discard it as non-exploitable. 
Second, dynamic elements often do not impact the leak methods 
(e.g., events fired, properties read).

\paragraph{Timing.}
\Tool supports the timing \xsleak through the video parsing technique
described in~\cite{VanGoethemClock2015CCS}. 
However, we did not use the timing \xsleak in our experiments, 
which may have prevented \tool from finding further attacks.
The main reason for disabling the timing \xsleak is that in order to 
attain the same level of reliability as other attack classes, 
it requires sending hundreds~\cite{VanGoethemClock2015CCS}, 
or even thousands~\cite{TimingML2019}, of HTTP requests per \sURL. 
This increases the load at the target and causes some web sites 
to respond with defenses (e.g., CAPTCHAs, blocking) that hamper the testing.
We noticed this initially on \url{linkedin.com}. 
In addition to the high load, we observed another three challenges 
in using the timing \xsleak.
First, we cannot generalize a timing attack. 
With timing, we always need to measure the timing for each URL in the
target site;
we cannot reuse what we learn from one attack in new attacks.
Second, timing information is harder to use as the number of states 
increases.
For example, if a URL allows downloading a file only to its owner,
there may not be a clear timing difference between an unauthenticated user and
an authenticated one that is not the owner.
Finally, it is hard to combine in the same attack 
timing with the non-timing \xsleaks.
Due to these challenges by default \tool does not use the timing \xsleak.
We leave applying timing leaks to more than two states for future work.

\paragraph{Discovering new \xsleaks.}
We have systematically explored existing COSI attacks and the \xsleaks 
they use, generalizing them into COSI attack classes. 
In this process, we have discovered a novel \textit{postMessage} \xsleak.
However, it is very likely that there exist more, currently unknown,  
\xsleaks leveraging other browser APIs. 
Systematically exploring the browser API surface to identify 
all possible \xsleaks remains an open challenge, 
which we plan to explore in future work.

\begin{table}
	\caption{Summary of previously proposed COSI attacks} 
	\label{tab:attacks}
	\centering
	\scriptsize
	\renewcommand{\arraystretch}{1.5}
	\begin{tabular}{lllll}\toprule
		
		\textbf{Reference} & \textbf{Year} & \textbf{Type} & \textbf{Attack Classes} &  \textbf{Browsers} \\ \midrule
		
		\multicolumn{1}{p{1.3cm}}{Grossman \& Hansen~\cite{GrossmanCosiImage2006}} & 2006 & Blog & EF-CtMismatchImg & - \\
		\multicolumn{1}{p{1.3cm}}{Grossman~\cite{Grossmancosijserror2006}} & 2006 & Blog &  JSError & F\\
		\multicolumn{1}{p{1.3cm}}{Shiflett~\cite{Shiflettcosijserror2006}} & 2006 & Blog &  JSError & F\\
		\multicolumn{1}{p{1.7cm}}{Bortz et al.~\cite{BortzTiming2007WWW}} & 2007 & Paper &  Timing & F, S\\
		\multicolumn{1}{p{1.3cm}}{Grossman~\cite{Grossmancold2008whoseproblem}} & 2008 & Blog &  \multicolumn{1}{p{2.4cm}}{EF-CtMismatchScript, EF-CtMismatchImg} & F\\
		\multicolumn{1}{p{1.3cm}}{Evans~\cite{Evanscsslo2008}} & 2008 & Blog &  CSSPropRead & F\\
		\multicolumn{1}{p{1.3cm}}{Evans~\cite{Evanssearchtiming2009}} & 2009 & Blog &  Timing & -\\
		\multicolumn{1}{p{1.3cm}}{Cardwell~\cite{CardwellColdScode2011}} & 2011 & Blog &  \multicolumn{1}{p{2.4cm}}{EF-StatusErrorScript, EF-CtMismatchImg} & C, F, IE\\
		\multicolumn{1}{p{1.3cm}}{Grossman~\cite{Grossmancoldmany2012}} & 2012 & Blog &  \multicolumn{1}{p{2.3cm}}{EF-StatusErrorIFrame, EF-CtMismatchScript, OP-LinkSheet, OP-FrameCount, EF-CtMismatchImg, JSObjectRead} & F \\
		\multicolumn{1}{p{1.3cm}}{Homakov~\cite{HomakovCSPbug2013}}  & 2013 & Bug & CSPViolation & C, F, IE\\
		\multicolumn{1}{p{1.3cm}}{Gelernter \& Herzberg~\cite{GelernterCSSA2015CCS}} & 2015 & Paper &  Timing & - \\
		\multicolumn{1}{p{1.9cm}}{Goethem et al.~\cite{VanGoethemClock2015CCS}}  & 2015 & Paper & \multicolumn{1}{p{1.3cm}}{Timing, EF-CtMismatchVideo} & C\\
		\multicolumn{1}{p{1.7cm}}{Lekies et al.~\cite{LekiesXSSI2015SEC}} & 2015 & Paper &  JSObjectRead & C\\
		\multicolumn{1}{p{1.6cm}}{Lee et al.~\cite{LeeAppCacheErr2015NDSS}} & 2015 & Paper &  AppCacheError & C\\
		\multicolumn{1}{p{1.9cm}}{Schwenk et al.~\cite{JorgSOPEval2017SEC}} & 2017 & Paper & OP-LinkSheet & IE, E\\
		\multicolumn{1}{p{1.3cm}}{Masas~\cite{Masasimpervaframelengthbug2018}} & 2018 & Blog & OP-WindowProperties & C\\
		\multicolumn{1}{p{1.3cm}}{Yoneuchi~\cite{Cspfingerprinting2018}} & 2018 & Blog & CSPViolation & F\\
		\multicolumn{1}{p{1.7cm}}{Gulyas et al.~\cite{GulyasCSPLO2018WPESextend}} &  2018 & Paper & CSPViolation & C\\
		\multicolumn{1}{p{1.3cm}}{Acar~\cite{AcarMediaError2018WAD}} & 2018 & Paper & OP-MediaStatus & C, F\\
		\multicolumn{1}{p{2.1cm}}{Staicu \& Pradel~\cite{StaicuLeakyImg2019SEC}} &  2019 & Paper & EF-CtMismatchImg & C, F\\
		\multicolumn{1}{p{1.3cm}}{Masas~\cite{MasasImpervaWLStat2019}} &  2019 & Blog & OP-WindowProperties & C\\
		\multicolumn{1}{p{1.9cm}}{Sanchez et al.~\cite{SanchezBakingTimer2019ACSAC}} &  2019 & Paper & Timing & C\\
		\multicolumn{1}{p{1.3cm}}{XSLeaks~\cite{XsleaksCollection2019}} & 2019 & Project & \multicolumn{1}{p{2.4cm}}{EF-CtMismatchImg, OP-FrameCount, CSPViolation, Timing, EF-CtMismatchObject, OP-ImgDimension, OP-MediaDuration, OP-WindowProperties, EF-CacheLoadCheck} & C, F, E\\
		\bottomrule
	\end{tabular}
\newline (\textbf{Legend:} F=Firefox; S=Safari; C=Chrome; IE=Internet Explorer; E=Edge; -= we couldn't find a browser mentioned in the article)
\vspace{-6pt}
\end{table}

\section{Related Work}
\label{sec:related}

\paragraph{Prior COSI attack instances.}
Table~\ref{tab:attacks} summarizes the \numPriorWorks prior works 
proposing COSI attack instances we have identified. 
The first instance of a COSI attack was proposed in 2006 by 
Grossman and Hansen~\cite{GrossmanCosiImage2006}. 
It was a login detection attack using the \textit{img} tag and 
the EventsFired \xsleak (\textit{EF-CtMismatchImg} attack class).
Since then, EventFired attacks have been shown to apply to other HTML tags 
and content types
\cite{CardwellColdScode2011,Grossmancold2008whoseproblem,Grossmancoldmany2012,XsleaksCollection2019}. 
Recently, Staicu and Pradel~\cite{StaicuLeakyImg2019SEC} showed that 
EventsFired attacks can be combined with shareable images 
to deanonymize users of image sharing services. 

In another blog post in 2006, Grossman~\cite{Grossmancosijserror2006}
introduced the first instance of the \textit{JSError} attack class 
that leverages the type and line number of errors triggered when a JavaScript 
resource is included using the \textit{script} tag.
This attack was then demonstrated on popular sites 
like Amazon~\cite{Shiflettcosijserror2006}. 
Inspired by Grossman's attacks, Evans~\cite{Evanscsslo2008} presented the 
first instance of the \textit{CSSPropRead} attack class, 
leveraging the presence of 
certain objects and variables from an included JS resource. 
In a 2012 post Grossman presented multiple attack instances 
including the first instances of the \textit{JSObjectRead} attack class 
and the first attack using the readable object properties \xsleak \cite{Grossmancoldmany2012}.
Lekies et al.~\cite{LekiesXSSI2015SEC} extended the \textit{JSObjectRead} 
class with more techniques such as prototype tampering 
and showed that \textit{JSObjectRead} attacks can be defended by making 
the URLs of script files unpredictable and including JS parser-breaking strings 
in dynamic JS files.
After Grossman's initial attack using the FrameCount readable object property, 
instances of attack classes leveraging 
other properties (e.g., window frame count, width, height, duration, cssRules, media error) 
have been proposed~\cite{Masasimpervaframelengthbug2018,MasasImpervaWLStat2019,XsleaksCollection2019,JorgSOPEval2017SEC,AcarMediaError2018WAD}.

Homakov~\cite{HomakovCSPbug2013,Homakovcsp4evil2014} showed that 
cross-origin and sub-domain redirections can be detected by abusing CSP. 
This approach has been used for login detection and fingerprinting 
attacks~\cite{GulyasCSPLO2018WPESextend,Cspfingerprinting2018}. 
Lee et al. showed that the AppCache feature can be abused to differentiate
between 200 status responses and redirection or error 
responses~\cite{LeeAppCacheErr2015NDSS}. 
Recently, Staicu et al. \cite{StaicuLeakyImg2019SEC} showed that a 
deanonymization attack can be mounted using images uploaded to GitHub. 
We generalized this attack on GitHub also to non-image resources.
Bortz et al.~\cite{BortzTiming2007WWW} 
showed that the timing of the events fired when a resource is loaded 
using the \textit{img} HTML tag is a 
good metric to determine the state of a user at a target site. 
Evans~\cite{Evanssearchtiming2009} and  
Gelernter and Herzberg~\cite{GelernterCSSA2015CCS} 
applied similar approaches for mounting cross-site search attacks. 
Goethem et al.~\cite{VanGoethemClock2015CCS} showed that the parsing time 
of the included resources is a better alternative and that 
the \texttt{Referer} and \texttt{Origin} headers
can help preventing such attacks. 
Recently, Sanchez et al.~\cite{SanchezBakingTimer2019ACSAC} 
have measured the scale of timing-based login and access detection attacks. 

This work shows that the above are all instances of 
COSI attacks, and demonstrates how to build 
complex COSI attacks that handle more than two states and multiple browsers.

\paragraph{Browser history sniffing attacks.}
Multiple works have studied history sniffing attacks 
that use browser side channels to determine whether a user has 
accessed certain web sites~\cite{FeltenTimingCacheDNS2000CCS,Clover2002css,WondracekSocialGroupDeanonym2010SnP,OlejnikJohny2012Hotpets,SmithHistoryRevisit2018WOOT}.
To defend against history sniffing attacks 
Jackson et al. proposed to increase the isolation of 
different origins~\cite{JacksonVisitedCaching2006WWW}
and Wondracek et al. proposed adding non-predictable tokens in URLs and 
using the POST method~\cite{WondracekSocialGroupDeanonym2010SnP}.
History sniffing attacks are similar to COSI attacks in leveraging a browser 
side channel, but fundamentally differ in the absence of a target site 
and in that the attack page does not send cross-origin requests.

\paragraph{Attacks using postMessage.}
Guan et al.~\cite{GuanPostMessPriv2016AsiaCCS} analyzed privacy issues 
in postMessages broadcasted by popular web sites and  
Stock et al. showed that usage of broadcasted postMessages 
has been increasing~\cite{StockTangled2017SEC}. 
Our postMessage \xsleak leverages differences between broadcasted 
postMessages in \sURLs and does not require that messages contain 
sensitive data.
\section{Conclusion}
\label{sec:conclusion}

We have presented COSI attacks as a comprehensive category and 
have introduced a novel approach to identify and build complex
COSI attacks that differentiate more than two states and
support multiple browsers.
Our approach combines multiple attack vectors,
possibly using different \xsleaks.
To enable our approach, we have introduced the concept of COSI attack classes 
and have proposed novel techniques to discover attack classes from existing 
instances of COSI attacks.
In this process, we have discovered a novel browser \xsleak based 
on {\it window.postMessage}.
We have implemented our approach into \tool,
a tool to find COSI attacks in a target web site.
We have applied \tool to test \nApps stand-alone web applications
and \nSites popular web sites,
finding COSI attacks against each of them.

\section*{Acknowledgments}
We thank Adam Doupe and the anonymous reviewers for their insightful
comments and feedback. 
This research was largely performed while Soheil Khodayari was an intern at
the IMDEA Software Institute.
This research has received funding from the European Union Horizon 2020 
Research and Innovation Programme under the ELASTEST Grant Agreement No. 731535.
This work was also supported by the Regional Government of Madrid 
through the BLOQUES-CM grant P2018/TCS-4339 
and by the Spanish Government through the SCUM grant RTI2018-102043-B-I00.
Any opinions, findings, and conclusions or recommendations expressed in 
this material are those of the authors or originators, and 
do not necessarily reflect the views of the sponsors.

\bibliographystyle{IEEEtranS}
\bibliography{bibliography}
\end{document}